\documentclass[%
reprint,
superscriptaddress,
amsmath,amssymb,
floatfix,
aps,
prl
]{revtex4-1}

\usepackage{graphicx}   %include figure files
\usepackage{dcolumn}% Align table columns on decimal point
\usepackage{bm}% bold math
\usepackage{hyperref}% add hypertext capabilities
\usepackage{epstopdf}
\usepackage{braket}
\usepackage{amsmath}
\usepackage{SIunits}
\usepackage{changes}
\usepackage[english]{babel}

\usepackage[colorinlistoftodos,prependcaption,textsize=tiny]{todonotes}
\usepackage[section]{placeins}

\makeatletter
\let\cat@comma@active\@empty
\makeatother
\newcommand{\qi}{Q_{\mathrm{i}}}

\newcommand{\be}{\begin{equation}}
\newcommand{\ee}{\end{equation}}
\newcommand{\ba}{\begin{equation}}
\newcommand{\ea}{\end{equation}}
\newcommand{\bea}{\begin{eqnarray}}
\newcommand{\eea}{\end{eqnarray}}

\newcommand{\eref}[1]{Eq.~(\ref{#1})}
\newcommand{\esref}[1]{Eqs.~(\ref{#1})}
\newcommand{\rref}[1]{(\ref{#1})}

\begin{document}

\title{Quasiparticle dynamics in granular aluminum \\ close to the superconductor to insulator transition}

\author{Lukas Gr\"unhaupt}
\affiliation{Physikalisches~Institut,~Karlsruhe~Institute~of~Technology,~76131~Karlsruhe,~Germany}

\author{Nataliya Maleeva}
\affiliation{Physikalisches~Institut,~Karlsruhe~Institute~of~Technology,~76131~Karlsruhe,~Germany}

\author{Sebastian T. Skacel}
\affiliation{Physikalisches~Institut,~Karlsruhe~Institute~of~Technology,~76131~Karlsruhe,~Germany}

\author{Martino Calvo}
\affiliation{Universit\'e~Grenoble~Alpes,~Institut~N\'eel,~F-38000 Grenoble,~France}
\affiliation{CNRS,~Institut~N\'eel,~F-38000~Grenoble,~France}

\author{Florence Levy-Bertrand}
\affiliation{Universit\'e~Grenoble~Alpes,~Institut~N\'eel,~F-38000 Grenoble,~France}
\affiliation{CNRS,~Institut~N\'eel,~F-38000~Grenoble,~France}

\author{Alexey V. Ustinov}
\affiliation{Physikalisches~Institut,~Karlsruhe~Institute~of~Technology,~76131~Karlsruhe,~Germany}
\affiliation{Russian Quantum Center, National University of Science and Technology MISIS, 119049 Moscow, Russia}

\author{Hannes Rotzinger}
\affiliation{Physikalisches~Institut,~Karlsruhe~Institute~of~Technology,~76131~Karlsruhe,~Germany}

\author{Alessandro Monfardini}
\affiliation{Universit\'e~Grenoble~Alpes,~Institut~N\'eel,~F-38000 Grenoble,~France}
\affiliation{CNRS,~Institut~N\'eel,~F-38000~Grenoble,~France}

\author{Gianluigi Catelani}
\affiliation{JARA~Institute~for~Quantum~Information~(PGI-11),~Forschungszentrum~J\"ulich,~52425~J\"ulich,~Germany}

\author{Ioan M. Pop}
\email{ioan.pop@kit.edu}
\affiliation{Physikalisches~Institut,~Karlsruhe~Institute~of~Technology,~76131~Karlsruhe,~Germany}
\affiliation{Institute~of~Nanotechnology,~Karlsruhe~Institute~of~Technology,~76344~Eggenstein~Leopoldshafen,~Germany}

\date{\today}

%-----------------------------------------------------------------------------

\begin{abstract}
Superconducting high kinetic inductance elements constitute a valuable resource for quantum circuit design and millimeter-wave detection.
Granular aluminum (GrAl) in the superconducting regime is a particularly interesting material since it has already shown a kinetic inductance in the range of \unit{\nano\henry\per\Box} and its deposition is compatible with conventional Al/AlOx/Al Josephson junction fabrication.
We characterize microwave resonators fabricated from GrAl with a room temperature resistivity of $\unit{4 \times 10^3}{\micro\ohm\cdot\centi\meter}$, which is a factor of 3 below the superconductor to insulator transition, showing a kinetic inductance fraction close to unity. 
The measured internal quality factors are on the order of $\qi = 10^5$ in the single photon regime, and we demonstrate that non-equilibrium quasiparticles (QP) constitute the dominant loss mechanism. 
We extract QP relaxation times in the range of \unit{1}{\second} and we observe QP bursts every \unit{\sim 20}{\second}. 
The current level of coherence of GrAl resonators makes them attractive for integration in quantum devices, while it also evidences the need to reduce the density of non-equilibrium QPs.
\end{abstract}

\maketitle

%%% INTRODUCTION %%%
Superconducting materials with a high kinetic inductance play a prominent role in superconducting circuits operating at microwave frequencies, such as quantum bits (qubits) with remarkably high energy relaxation times \cite{manucharyan_fluxonium_2009, pop_coherent_2014, lin_protecting_2017, ernest_realization_2017}, topological \cite{gladchenko_superconducting_2009, brooks_protected_2013} and protected qubits \cite{richer_inductively_2017, groszkowski_coherence_2017, petrescu_fluxon-based_2017}, cohe\-rent quantum phase slip circuits \cite{astafiev_coherent_2012, belkin_formation_2015, bell_spectroscopic_2016, muller_passive_2017}, wideband parametric amplifiers \cite{ho_eom_wideband_2012, vissers_low-noise_2016}, and resonators with custom designed Kerr non-linearity for quantum state of light engineering \cite{cohen_degeneracy-preserving_2017, puri_engineering_2017}. 
As the kinetic inductance fraction $\alpha = L_{\mathrm{kinetic}} / L_{\mathrm{total}}$ increases, so does the susceptibility of superconducting circuits to quasiparticle (QP) excitations (broken Cooper pairs), which constitutes an asset for kinetic inductance detectors (KIDs) \cite{day_broadband_2003}. 
In contrast, for quantum information applications, the circuits are heavily shielded, in an effort to minimize the generation of excess QPs, due to photons, phonons, or other particles with energies larger than twice the superconducting gap. 
Even residual QP densities as low as $10^{-6}$, norma\-lized to the density of Cooper pairs, can be responsible for excess decoherence in superconducting quantum circuits \cite{aumentado_nonequilibrium_2004, de_visser_number_2011, maisi_excitation_2013, levenson-falk_single-quasiparticle_2014, pop_coherent_2014, Wang_NatComm2014, bilmes_electronic_2017}. 
For temperatures much lower than the critical temperature, in the limit of weak microwave drive, the origin and dynamics of excess QPs is an active field of research \cite{Wang_NatComm2014, nsanzineza_trapping_2014, levenson-falk_single-quasiparticle_2014, janvier_coherent_2015, gustavsson_suppressing_2016, taupin_tunable_2016}, with direct implications for the implementation of quantum computation with Majorana modes \cite{higginbotham_parity_2015, aasen_milestones_2016}.

\begin{figure}[tb]
\centering
\includegraphics[width=1\columnwidth]{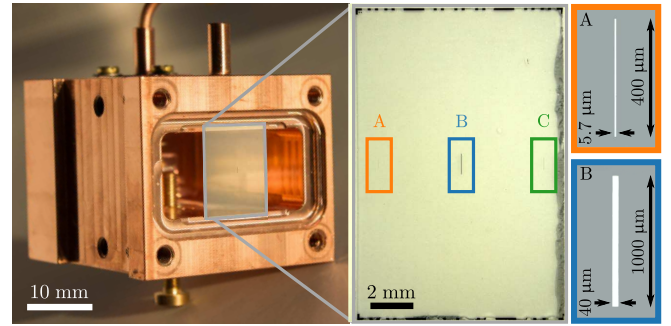}
\caption{\label{fig:1} Optical images of the GrAl stripline resonators and the 3D waveguide sample holder. The copper waveguide provides a low-loss environment \cite{zoepfl_characterization_2017, kou_simultaneous_2017} for three GrAl resonators patterned by e-beam lift-off lithography on a \unit{10 \times 15}{\milli\meter\squared} c-plane sapphire chip. The resonator dimensions are the following: A - \unit{400 \times 5.4}{\micro\meter\squared} (orange, left), B - \unit{1000 \times 40}{\micro\meter\squared} (blue, middle), and C - \unit{600 \times 10}{\micro\meter\squared} (green, right). The \unit{20}{\nano\meter} thick GrAl film has a resistivity $\rho = \unit{4 \times 10^3}{\micro\ohm\cdot\centi\meter}$, corresponding to a sheet resistance $R_{\mathrm{s}} = \unit{2}{\kilo\ohm\per\Box}$. By comparing the measured resonant frequencies (cf. Table~\ref{tab:1}) with a finite elements method (FEM) simulation we extract a kinetic inductance $L_{\mathrm{kinetic}} = \unit{2}{\nano\henry\per\Box}$.}
\end{figure}

\begin{figure*}[t]
\centering
\includegraphics[scale=1]{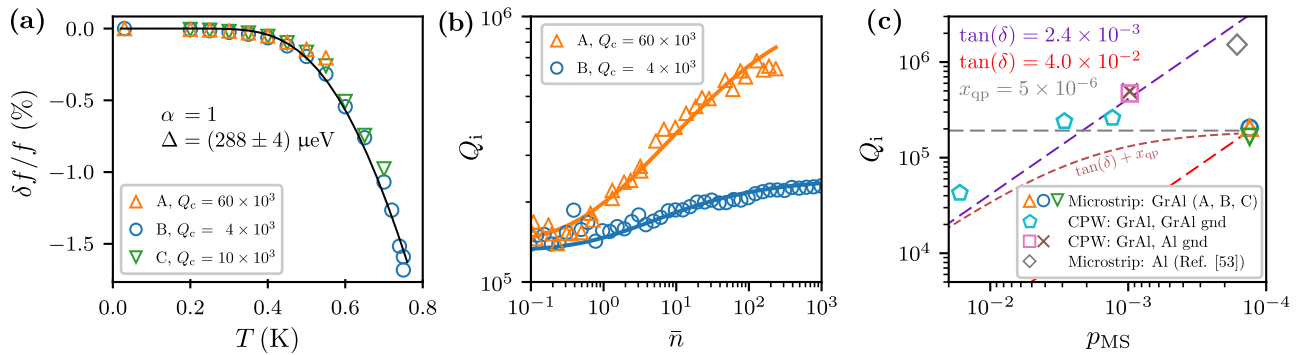}
\caption{\label{fig:2} Measurement of radio frequency loss mechanisms in GrAl films. \textbf{(a)} Measurement of the relative shift of the resonant frequency $\delta f / f = \left( f(T) - f(0.02) \right) / f(0.02)$ as a function of temperature. From FEM simulations we expect the kinetic inductance to be three orders of magnitude larger than the geometric inductance. The data can be fitted using a Bardeen-Cooper-Schrieffer (BCS) model \cite{gao_physics_2008, turneaure_surface_1991}, $\delta f(T)/f = -\frac{\alpha}{2} \sqrt{\pi \Delta / (2 k_{\mathrm{B}} T)} \exp(-\Delta / k_{\mathrm{B}} T)$, which is expected to approximately describe the temperature dependence of the frequency, but which does not take into account corrections to the prefactor due to deviations of GrAl from standard BCS theory \cite{pracht_enhanced_2016}, or due to the regime $k_\mathrm{B} T \sim hf$.  The black line shows the fit for the kinetic inductance fraction $\alpha = 1$. We extract a value for the GrAl superconducting gap $\Delta = \unit{(288 \pm 4)}{\micro\electronvolt}$, in agreement, within $15\%$, with the measured gap from THz spectroscopy \cite{pracht_enhanced_2016, dupre_phase_2017}, and previously reported values \cite{deutscher_transition_1973}. Notice that the GrAl gap is $ \sim 1.6$ times larger than that of thin film aluminum. When using $\alpha$ as a free fit parameter, we obtain similar values ($\alpha = 0.92 \pm 0.06$, $\Delta = \unit{(282 \pm 4)}{\micro\electronvolt}$). \textbf{(b)} Measured internal quality factors $Q_{\mathrm{i}}$ as a function of the average circulating photon number $\bar{n}$. The solid lines represent a fit to the QP activation model of Eq.~\eqref{eq:3}, discussed in the main text. \textbf{(c)} Comparison between measured $Q_{\mathrm{i}}$ in the single photon regime as a function of the metal-substrate participation ratio $p_{\mathrm{MS}}$ for different resonator geometries (see supplementary). These results suggest that high kinetic inductance GrAl resonators A-C are limited by excess QPs, not by surface dielectric loss. GrAl resonators in CPW geometry, with $p_{\mathrm{MS}} > 10^{-3}$, are limited by a surface dielectric loss tangent $\tan{(\delta)} = 2.4 \times 10^{-3}$, similar to aluminum qubits \cite{wang_surface_2015}. For a detailed discussion see the main text.}
\end{figure*}

In this letter we report the measurement of QP dynamics in microwave resonators fabricated from superconducting granular aluminum (GrAl) \cite{buckel_einfluss_1954, cohen_superconductivity_1968, deutscher_transition_1973, pracht_enhanced_2016}, with a kinetic inductance fraction close to unity \cite{rotzinger_aluminium-oxide_2017}. 
At a temperature $ T = \unit{25}{\milli\kelvin}$, orders of magnitude below the criti\-cal temperature $T_\mathrm{c} = \unit{2.1}{\kelvin}$, we observe QP bursts every $\sim \unit{20}{\second}$ and QP relaxation times $\tau_{\mathrm{ss}}$ in the range of $\unit{1}{\second}$. 
Despite such remarkably long relaxation times, we show that GrAl resonators with a kinetic inductance $L_{\mathrm{kinetic}}$ as high as \unit{2}{\nano\henry\per\Box} maintain internal quality factors $\qi$ in the range of $10^5$ for $\bar{n} \approx 1$ average circula\-ting photons. 
These properties place GrAl in the same class of low-loss, high kinetic inductance environments as Josephson junction arrays \cite{manucharyan_superinductance_2012, masluk_microwave_2012, bell_quantum_2012} and disordered superconducting thin films, such as TiN \cite{vissers_low_2010, leduc_titanium_2010, swenson_operation_2013}, NbTiN \cite{barends_reduced_2010, samkharadze_high_kinetic_inductance_2016}, and NbN \cite{grabovskij_textitsitu_2008, luomahaara_kinetic_2014}.

Granular aluminum films owe their name to the self-assembly of pure aluminum grains, $\sim \unit{3}{\nano\meter}$ in diameter \cite{cohen_superconductivity_1968, deutscher_transition_1973}, inside an aluminum oxide matrix, during the deposition of pure aluminum in an oxygen atmosphere ($p_{\mathrm{Ox}} \sim \unit{10^{-5}}{\milli\bbar}$). 
Controlling the oxygen pressure allows the fabrication of films with resistivities $\rho$ in the range of  $1 - \unit{10^5}{\micro\ohm\cdot\centi\meter}$. 
The superconducting critical temperature $T_{\mathrm{c}}$ increases with $\rho$ to a maximum value of about $\unit{2.2}{\kelvin}$ at $\rho = \unit{4 \times 10^2}{\micro\ohm\cdot\centi\meter}$, and GrAl undergoes a superconductor to insulator transition (SIT) at $\rho \approx \unit{10^4}{\micro\ohm\cdot\centi\meter}$ \cite{pracht_enhanced_2016, dupre_phase_2017}. 
The main results in this letter are obtained for stripline resonators (see Fig.~\ref{fig:1}) fabricated from a GrAl film with $\rho = \unit{4 \times 10^3}{\micro\ohm\cdot\centi\meter}$, and a corresponding $T_{\mathrm{c}} = \unit{2.1}{\kelvin}$. 
The resistivity was choosen as high as possible, to maximize the kinetic inductance $L_{\mathrm{kinetic}} \propto \rho$ \cite{rotzinger_aluminium-oxide_2017}, while remaining sufficiently below the SIT, where quantum fluctuations and film inhomogeniety start dominating the microwave properties \cite{feigelman_microwave_2018}.  

%%% SAMPLE FABRICATION %%%

Figure~\ref{fig:1} shows a photograph of a 3D copper wave\-guide sample holder, which provides a well controlled, low loss microwave environment. 
Furthermore, the wave\-guide design dilutes the electric field strength, thereby reducing the surface dielectric participation ratio \cite{zoepfl_characterization_2017, kou_simultaneous_2017}. 
The $\unit{20}{\nano\meter}$ thick GrAl resonators are patterned on a $\unit{10 \times 15}{\milli\meter\squared}$ c-plane sapphire chip, using e-beam lift-off lithography, on a PMMA/MMA bilayer. 
By varying the length of the resonators, and therefore their electric dipole moment, we tune the coupling quality factor $Q_{\mathrm{c}}$ from $4 \times 10^3$ to $6 \times 10^4$ (see Table~\ref{tab:1}). 

\begin{table}[bht]
\begin{center}
\caption{\label{tab:1} Dimensions, coupling quality factors, and resonant frequencies of resonators A, B, and C. Between cool downs (run \#1, \#2, \#3) the samples were stored in atmosphere at room temperature for two, and four weeks respectively.}
\begin{tabular*}{1\columnwidth}{@{\extracolsep{\fill}}crrccc}
	\hline \hline
	 & \multicolumn{1}{c}{dimensions} & \multicolumn{1}{c}{$Q_{\mathrm{c}}$} & \multicolumn{3}{c}{resonant frequency (GHz)} \\
	 & \multicolumn{1}{c}{(\unit{\micro\meter\squared})} & $(10^3)$ & run \#1 & run \#2 & run \#3 \\
	\hline
	A & $400 \times\,~5.4$ & $60$  & 7.006 & 6.999 & 6.994 \\
	\hline
	B& $1000 \times 40.0$ & $4$  & 6.032 & 6.027 & 6.025 \\
	\hline
	C & $600 \times 10.0$  & $10$  & 6.330 & 6.322 & 6.287 \\ 
	\hline \hline 
\end{tabular*}
\end{center}
\end{table}

\begin{figure*}[tb]
\centering
\def\svgwidth{\columnwidth}
\includegraphics[scale=1]{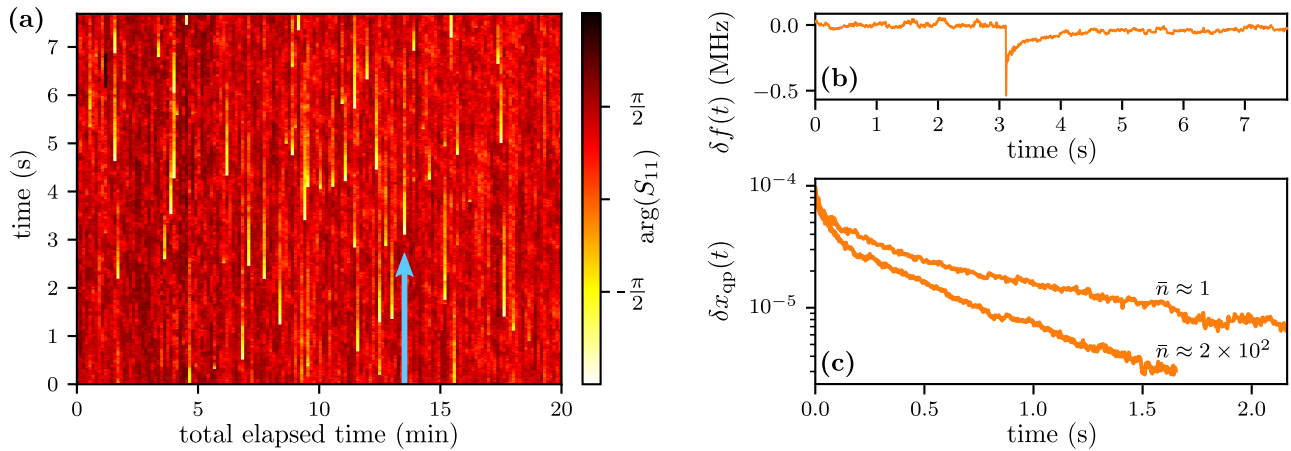}
\caption{\label{fig:3} Measurement of QP generating events. \textbf{(a)} Typical plot of a continuous monitoring of a resonator's phase signal at one frequency point. Multiple time traces are consecutively recorded, covering a total time of about \unit{45}{\minute} (for clarity only partially shown). The measurement reveals discrete jumps of the resonant frequency to a lower value followed by a relaxation over seconds approximately every \unit{20}{\second}.
\textbf{(b)} Plot of the time trace indicated by the arrow in panel (a), where the phase response is converted into a frequency shift, showing an instantaneous drop followed by a slow relaxation. \textbf{(c)} By recording multiple events and averaging them (see supplementary), an exponential tail can be seen. The characteristic relaxation time, $\tau_{\mathrm{ss}}$, depends on the average circulating photons $\bar{n}$.}
\end{figure*}

We perform standard microwave reflection measurements in commercial dilution cryostats with (run \#1) and without a liquid helium bath (runs \#2 and \#3), at a base temperature $T \approx \unit{25}{\milli\kelvin}$. 
The waveguide sample holders are successively surrounded by a series of shields and absorbing materials, to minimize stray radiation and magnetic fields (cf. \cite{grunhaupt_argon_2017} and supplementary).

%%% RESULTS DISCUSSION %%%

%Discussion Fig. 2a

Comparing the measured resonant frequencies of the GrAl resonators with FEM simulations, we infer a kinetic inductance $L_{\mathrm{kinetic}} = \unit{2}{\nano\henry\per\Box}$ for $\rho = \unit{4 \times 10^3}{\micro\ohm\cdot\centi\meter}$. 
The corresponding kinetic inductance fraction $\alpha$ is expected to be close to unity, as confirmed by measurements in Fig.~\ref{fig:2}a.
For superconducting quantum circuits with $\alpha = 1$, internal quality factors $\qi$ could start to be limited by residual QPs. 
We measure single photon $\qi$ on the order of $10^5$ (see Fig.~\ref{fig:2}b), comparable with other realizations of high kinetic inductance materials \cite{weisl_kerr_2015, masluk_microwave_2012, vissers_low_2010, samkharadze_high_kinetic_inductance_2016}, which could be explained by a residual excess quasiparticle density $x_{\mathrm{qp}} = 5 \times 10^{-6}$, in the range of previously reported values \cite{aumentado_nonequilibrium_2004,de_visser_number_2011,maisi_excitation_2013,levenson-falk_single-quasiparticle_2014,pop_coherent_2014,Wang_NatComm2014,bilmes_electronic_2017}.

%Discussion Fig. 2b
Figure~\ref{fig:2}b also shows the $\qi$ dependence on the average circulating photon number $\bar{n} =  4 P_{\mathrm{in}} Q_{\mathrm{tot}}^2 / \left(\hbar \omega_{\mathrm{r}}^2 Q_{\mathrm{c}}\right)$. 
For resonator A, $\qi$ shows an increase by a factor of four between $\bar{n} = 1$ and $100$, reaching $6 \times 10^5$ before the re\-sonator bifurcates due to its intrinsic non-linearity \cite{maleeva_circuit_2018}. 
Seven times larger in cross section, resonator B shows a smaller increase in $\qi$, which is less than a factor of two for $\bar{n}$ between $1$ and $10^4$. 
The internal quality factor of resonator C could not be fitted above $\bar{n} \approx 1$ because the amplitude data shows an irregular behavior, changing from the expected dip to a peak (see supplementary).  
This may be caused, among other reasons, by impedance imperfections in the measurement setup, or local flux trapping.

The measured increase of $\qi$ with $\bar{n}$ can be attributed to the saturation of dielectric loss \cite{hunklinger_saturation_1972, golding_nonlinear_1973}, or the activation of QPs \cite{gustavsson_suppressing_2016, levenson-falk_single-quasiparticle_2014}. 
The measurements summarized in Fig.~\ref{fig:2}c offer additional insight into the dominant loss mechanism for resonators A-C. 
As we will discuss in the following, they indicate that $\qi$ is QP limited, and consequently suggest a small contribution from dielectric loss saturation to the $\qi$ power dependence. 

Concretely, Fig.~\ref{fig:2}c shows a comparison of measured single photon $\qi$ for different GrAl resonator geometries as a function of their metal-substrate interface partici\-pation ratio $p_{\mathrm{MS}}$, following the methodology in Refs. \cite{wenner_surface_2011, wang_surface_2015}. 
The results for resonators A-C, which are the main focus of the letter, are shown by triangles and a circle, and they are about a factor of ten lower than the typically measured dielectric loss tangent $\tan(\delta)  = 2.6 \times 10^{-3}$ \cite{wang_surface_2015}. This could either be explained by a ten times larger dielectric loss tangent associated to the GrAl film or by the presence of an additional loss mechanism, such as QP dissipation. 
In order to distinguish between these two possibilities, we measured single photon $\qi$ for GrAl samples with increasingly larger $p_{\mathrm{MS}}$. 
To perform these control experiments, we employed coplanar waveguide (CPW) resonator designs, which are convenient to increase the $p_{\mathrm{MS}}$. 
The blue pentagons in Fig.~\ref{fig:2}c show that increasing the $p_{\mathrm{MS}}$ by a factor of ten does not degrade the quality factor. 
To observe a decrease in $\qi$ by a factor of five, we had to increase the $p_{\mathrm{MS}}$ by two orders of magnitude compared to samples A-C, indica\-ting that dielectric loss is not dominant. 
Surprinsingly, when the ground plane is fabricated from aluminum, we observe an increase of $\qi$ by a factor of two for an increase of the $p_{\mathrm{MS}}$ by a factor of ten, presumably due to phonon trapping in the lower gap aluminum ground plane \cite{daddabbo_applications_2014}. This result directly suggests QPs as dominant dissipation source, which is confirmed by measurements on aluminum resonators \cite{grunhaupt_argon_2017}, with ten times smaller $\alpha$ compared to resonators A-C, and similar $p_{\mathrm{MS}}$, showing approximately a factor of ten increase in $\qi$ (cf. grey rhombus in Fig.~\ref{fig:2}c).

\begin{figure}[thb]
\centering
\def\svgwidth{\columnwidth}
\includegraphics[scale=1]{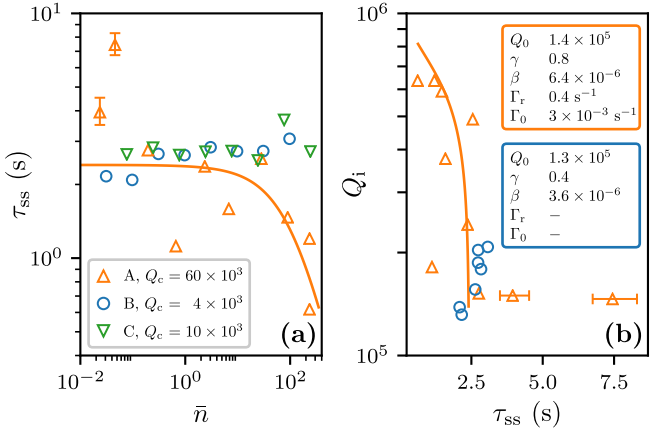}
\caption{\label{fig:4} Steady state QP relaxation constant $\tau_{\mathrm{ss}}$ as a function of $\bar{n}$, and its correlation with $\qi$. \textbf{(a)} Resonator A (orange triangles) shows a decrease of $\tau_{\mathrm{ss}}$ by approximately an order of magnitude whereas $\tau_{\mathrm{ss}}$ of B and C stays constant, within a factor of two. The solid line is a fit to the phenomenological model described by Eq.~\eqref{eq:4}. All fit parameters are given in the inset of panel (b). Error bars, where not plotted, are approximately the size of the marker and represent the statistical error of the fit (see supplementary). \textbf{(b)} Correlation between $\tau_{\mathrm{ss}}$ and $Q_{\mathrm{i}}$ for resonators A and B. The orange line is traced using the fits for the measured $\tau_{\mathrm{ss}}$ (panel (a)) and $\qi$ (Fig.~\ref{fig:2}b).}
\end{figure}

A possible source of excess QPs could be impacts of high energy particles, documented in the KID community \cite{swenson_high-speed_2010, moore_position_2012, cardani_energy_2015}. 
By continuously monitoring the phase response of the resonators, we observe sudden drops of the resonant frequency, appearing stochastically every $\sim \unit{20}{\second}$ (see supplementary), as shown in Fig.~\ref{fig:3}a, followed by a remarkably long relaxation process on the timescale of seconds (cf. Fig.~\ref{fig:3}b). 
Following a QP burst, the resulting QP density change, $\delta x_{\mathrm{qp}}(t)$, decreases the superfluid density, thereby increasing the kinetic inductance and lowering the resonant frequency $f$ by $2 \, \delta f(t) \propto - f \alpha \, \delta x_{\mathrm{qp}}(t)$ \cite{Wang_NatComm2014}.
 
Fig.~\ref{fig:3}c shows the relaxation of  $\delta x_{\mathrm{qp}}(t)$ following a QP burst, obtained from an average over tens of individual events, over the course of $\unit{45}{\minute}$ (see supplementary). Notice that the QP relaxation only becomes exponential at long timescales where it is dominated by single QP relaxation, with characteristic timescale $\tau_{\mathrm{ss}}$, whereas at short times the process is governed by QP recombination, as previously evidenced in superconducting qubits \cite{Wang_NatComm2014}. 
Surprisingly, $\tau_{\mathrm{ss}}$ depends on $\bar{n}$, as shown by the two curves in Fig.~\ref{fig:3}c, changing from $\tau_{\mathrm{ss}} \approx \unit{0.6}{\second}$ at $\bar{n} = 300$, to $\tau_{\mathrm{ss}} \approx \unit{1 - 7}{\second}$ in the single photon regime (see Fig.~\ref{fig:4}a). 
The statistical error for each $\tau_{\mathrm{ss}}$ measurement is much smaller than the observed scattering between the points, possibly due to slow fluctuations in the background QP population \cite{pop_coherent_2014}.

%%% Discussion of Model %%%

The dependence of $\tau_{\mathrm{ss}}$ with $\bar{n}$ suggests that the circulating power in the resonator can accelerate QP diffusion, as reported in qubits \cite{gustavsson_suppressing_2016} and nanobridge junctions \cite{levenson-falk_single-quasiparticle_2014}. This behavior is different from the previously reported QP generation in aluminum resonators at large driving powers, where $\bar{n} \approx 10^6$~\cite{de_visser_evidence_2014}.
In the presence of disorder, there are spatial variations of the superconducting order parameter where QPs can be localized, giving rise to a small sub gap tail in the density of states (DoS) \cite{bespalov_theoretical_2016}. Similarly to Ref. \cite{rothwarf_measurement_1967}, we develop a phenomenological model that accounts for QP generation, QP recombination, and transition between localized, $x_l$, and mobile, $x_m$, QP densities:

\begin{eqnarray*}
\dot{x}_m & = & - \Gamma_{mm} x_m^2 - \Gamma_{ml} x_m x_l - \Gamma_\mathrm{loc} x_m + \Gamma_\mathrm{ex} x_l + g_m \label{
}, \\
\dot{x}_l & = & - \Gamma_{ll} x_l^2 - \Gamma_{ml} x_m x_l + \Gamma_\mathrm{loc} x_m - \Gamma_\mathrm{ex} x_l + g_l \label{xleq}. 
\end{eqnarray*}

In our model, $\Gamma_{mm}$, $\Gamma_{ll}$, and $\Gamma_{ml}$ are rates at which two mobile, two localized, or a mobile and a localized QP recombine. 
$\Gamma_\mathrm{loc}$ and $\Gamma_\mathrm{ex}$ account for QP localization and excitation (due to photon interaction) from states in the sub gap tail of the DoS, while $g_m$ and $g_l$ describe the generation of mobile and localized QPs. The photon assisted excitation of localized QPs can be modeled by $\Gamma_\mathrm{ex} = \Gamma_0 \bar{n}$, where $\Gamma_0$ accounts for the strength of photon interaction with QPs. In principle, both mobile and localized QPs contribute to loss, proportional to their normalized density, coupling strength, and final DoS. Assuming a larger final DoS for localized QPs, and further simplifications (see supplementary), we model the photon number dependent loss due to localized QPs,
\begin{equation}
\frac{1}{Q_i} = \frac{1}{Q_0} + \beta \left[ \frac{1}{1+\frac{\gamma\bar{n}}{1+ \frac12\left(\sqrt{1+4\gamma\bar{n}}-1\right)}} - 1\right]. 
\label{eq:3}
\end{equation} 
Here, $1/Q_0$ quantifies residual loss mechanisms independent of $\bar{n}$, $\gamma = 2 \Gamma_\mathrm{loc} \Gamma_0 / (g_m \Gamma_{ml})$, and $\beta \propto \Gamma_\mathrm{loc} / \Gamma_{ml}$, but also accounts for QP-photon coupling strength, and final DoS. 
Similarly, we model $1 / \tau_{\mathrm{ss}}$ by a residual decay rate $\Gamma_r$ and a photon dependent part, 
\begin{equation}
\frac{1}{\tau_\mathrm{ss}} = \Gamma_{r} + \Gamma_0 \left[\bar{n} + \frac{1}{2\gamma}\left(\sqrt{1+4\gamma\bar{n}}-1\right)\right],
\label{eq:4}
\end{equation}
which indicates that QPs relax faster at higher photon numbers $\bar{n}$, when they are activated from localized to mobile.

The fit parameters for resonators A and B are given in the insets in Fig.~\ref{fig:4}b. As expected from Fig.~\ref{fig:2}b, both resonators show comparable residual quality factors, $Q_0 \approx \qi \, (\bar{n} \ll 1)$.  
The ratio between the $\beta$ coefficients for resonators A and B is $\sim 2$, comparable to the ratio between the $\gamma$ coefficients. Since both $\beta$ and $\gamma$ are proportional to the QP-photon coupling (see supplementary), the factor of two difference between resonator A and B might be explained by the reduced current density per photon in resonator B, due to its seven times wider cross section. 
The flat behavior of $\tau_{ss}$ for resonator B suggests that the QP-photon coupling is not sufficient to excite localized QPs. 
Consequently, the small increase in $\qi$ as a function of $\bar{n}$ for resonator B might be explained by the saturation of dielectric loss, which is a secondary loss mechanism, accounting for approximately $10\%$ of the total loss (see. Fig.~\ref{fig:2}c). 
In contrast, $\tau_{\mathrm{ss}}(\bar{n})$ and $\qi(\bar{n})$ for resonator A appear to be correlated (see Fig.~\ref{fig:4}b), and can be fitted by our phenomenological model.

%%%% CONCLUSION %%%
In summary, we characterized GrAl superconduc\-ting microwave resonators with a kinetic inductance $L_{\mathrm{kinetic}}~=~\unit{2}{\nano\henry\per\Box}$, and internal quality factors $\qi > 10^5$ in the single photon regime, dominated by dissipation due to non-equilibrium QPs.
A continuous monitoring of the resonant frequencies shows stochastic QP bursts every $\sim \unit{20}{\second}$, followed by an exceptionally long QP relaxation time in the range of seconds, several orders of magnitude longer than in aluminum films \cite{de_visser_evidence_2014, Wang_NatComm2014, swenson_high-speed_2010}, or in Josephson junction superinductances \cite{vool_non-poissonian_2014}, presumably explained by single QP localization in regions of lower gap.

Interestingly, despite the disordered nature of the aluminum oxide in-between the aluminum grains in GrAl, we measure a similar dielectric loss tangent compared to previously reported values in a variety of pure aluminum superconducting quantum circuits \cite{wang_surface_2015}. The coherence properties of GrAl resonators are promising for high impedance superconducting quantum circuits and for ultra-sensitive KIDs. However, to be able to harvest the full potential of GrAl high kinetic inductance films, the density of excess QPs needs to be further reduced, either by elucidating the origin of the QP bursts, which cannot be inferred from our data, or by the use of phonon and QP traps \cite{riwar_normal-metal_2016}.

%%% Acknowledgement %%%
\begin{acknowledgments}
We are grateful to K. Serniak, M. Hays, and M. H. Devoret for insightful discussions, and to L. Radtke, A. Lukashenko, F. Valenti, and P. Winkel for technical support. Facilities use was supported by the KIT Nanostructure Service Laboratory (NSL). 
Funding was provided by the Alexander von Humboldt foundation in the framework of a Sofja Kovalevskaja award endowed by the German Federal Ministry of Education and Research, and by the Initiative and Networking Fund of the Helmholtz Association, within the Helmholtz Future Project \textit{Scalable solid state quantum computing}.
This work was partially supported by the Ministry of Education and Science of the Russian Federation in the framework of the Program to Increase Competitiveness of the NUST MISIS, contracts no. K2-2016-063 and K2-2017-081.
\end{acknowledgments}

\bibliography{qp_dynamics_GrAl_abbr}

\begin{thebibliography}{68}%
\makeatletter
\providecommand \@ifxundefined [1]{%
 \@ifx{#1\undefined}
}%
\providecommand \@ifnum [1]{%
 \ifnum #1\expandafter \@firstoftwo
 \else \expandafter \@secondoftwo
 \fi
}%
\providecommand \@ifx [1]{%
 \ifx #1\expandafter \@firstoftwo
 \else \expandafter \@secondoftwo
 \fi
}%
\providecommand \natexlab [1]{#1}%
\providecommand \enquote  [1]{``#1''}%
\providecommand \bibnamefont  [1]{#1}%
\providecommand \bibfnamefont [1]{#1}%
\providecommand \citenamefont [1]{#1}%
\providecommand \href@noop [0]{\@secondoftwo}%
\providecommand \href [0]{\begingroup \@sanitize@url \@href}%
\providecommand \@href[1]{\@@startlink{#1}\@@href}%
\providecommand \@@href[1]{\endgroup#1\@@endlink}%
\providecommand \@sanitize@url [0]{\catcode `\\12\catcode `\$12\catcode
  `\&12\catcode `\#12\catcode `\^12\catcode `\_12\catcode `\%12\relax}%
\providecommand \@@startlink[1]{}%
\providecommand \@@endlink[0]{}%
\providecommand \url  [0]{\begingroup\@sanitize@url \@url }%
\providecommand \@url [1]{\endgroup\@href {#1}{\urlprefix }}%
\providecommand \urlprefix  [0]{URL }%
\providecommand \Eprint [0]{\href }%
\providecommand \doibase [0]{http://dx.doi.org/}%
\providecommand \selectlanguage [0]{\@gobble}%
\providecommand \bibinfo  [0]{\@secondoftwo}%
\providecommand \bibfield  [0]{\@secondoftwo}%
\providecommand \translation [1]{[#1]}%
\providecommand \BibitemOpen [0]{}%
\providecommand \bibitemStop [0]{}%
\providecommand \bibitemNoStop [0]{.\EOS\space}%
\providecommand \EOS [0]{\spacefactor3000\relax}%
\providecommand \BibitemShut  [1]{\csname bibitem#1\endcsname}%
\let\auto@bib@innerbib\@empty
%</preamble>
\bibitem [{\citenamefont {Manucharyan}\ \emph {et~al.}(2009)\citenamefont
  {Manucharyan}, \citenamefont {Koch}, \citenamefont {Glazman},\ and\
  \citenamefont {Devoret}}]{manucharyan_fluxonium_2009}%
  \BibitemOpen
  \bibfield  {author} {\bibinfo {author} {\bibfnamefont {V.~E.}\ \bibnamefont
  {Manucharyan}}, \bibinfo {author} {\bibfnamefont {J.}~\bibnamefont {Koch}},
  \bibinfo {author} {\bibfnamefont {L.~I.}\ \bibnamefont {Glazman}}, \ and\
  \bibinfo {author} {\bibfnamefont {M.~H.}\ \bibnamefont {Devoret}},\ }\href
  {\doibase 10.1126/science.1175552} {\bibfield  {journal} {\bibinfo  {journal}
  {Science}\ }\textbf {\bibinfo {volume} {326}},\ \bibinfo {pages} {113}
  (\bibinfo {year} {2009})}\BibitemShut {NoStop}%
\bibitem [{\citenamefont {Pop}\ \emph {et~al.}(2014)\citenamefont {Pop},
  \citenamefont {Geerlings}, \citenamefont {Catelani}, \citenamefont
  {Schoelkopf}, \citenamefont {Glazman},\ and\ \citenamefont
  {Devoret}}]{pop_coherent_2014}%
  \BibitemOpen
  \bibfield  {author} {\bibinfo {author} {\bibfnamefont {I.~M.}\ \bibnamefont
  {Pop}}, \bibinfo {author} {\bibfnamefont {K.}~\bibnamefont {Geerlings}},
  \bibinfo {author} {\bibfnamefont {G.}~\bibnamefont {Catelani}}, \bibinfo
  {author} {\bibfnamefont {R.~J.}\ \bibnamefont {Schoelkopf}}, \bibinfo
  {author} {\bibfnamefont {L.~I.}\ \bibnamefont {Glazman}}, \ and\ \bibinfo
  {author} {\bibfnamefont {M.~H.}\ \bibnamefont {Devoret}},\ }\href {\doibase
  10.1038/nature13017} {\bibfield  {journal} {\bibinfo  {journal} {Nature}\
  }\textbf {\bibinfo {volume} {508}},\ \bibinfo {pages} {369} (\bibinfo {year}
  {2014})}\BibitemShut {NoStop}%
\bibitem [{\citenamefont {Lin}\ \emph {et~al.}(2017)\citenamefont {Lin},
  \citenamefont {Nguyen}, \citenamefont {Grabon}, \citenamefont {San~Miguel},
  \citenamefont {Pankratova},\ and\ \citenamefont
  {Manucharyan}}]{lin_protecting_2017}%
  \BibitemOpen
  \bibfield  {author} {\bibinfo {author} {\bibfnamefont {Y.-H.}\ \bibnamefont
  {Lin}}, \bibinfo {author} {\bibfnamefont {L.~B.}\ \bibnamefont {Nguyen}},
  \bibinfo {author} {\bibfnamefont {N.}~\bibnamefont {Grabon}}, \bibinfo
  {author} {\bibfnamefont {J.}~\bibnamefont {San~Miguel}}, \bibinfo {author}
  {\bibfnamefont {N.}~\bibnamefont {Pankratova}}, \ and\ \bibinfo {author}
  {\bibfnamefont {V.~E.}\ \bibnamefont {Manucharyan}},\ }\href
  {https://arxiv.org/abs/1705.07873} {\bibfield  {journal} {\bibinfo  {journal}
  {arXiv:1705.07873 [cond-mat.supr-con]}\ ,\ } (\bibinfo {year}
  {2017})}\BibitemShut {NoStop}%
\bibitem [{\citenamefont {Ernest}\ \emph {et~al.}(2017)\citenamefont {Ernest},
  \citenamefont {Chakram}, \citenamefont {Lu}, \citenamefont {Irons},
  \citenamefont {Naik}, \citenamefont {Leung}, \citenamefont {Lawrence},
  \citenamefont {Koch},\ and\ \citenamefont
  {Schuster}}]{ernest_realization_2017}%
  \BibitemOpen
  \bibfield  {author} {\bibinfo {author} {\bibfnamefont {N.}~\bibnamefont
  {Ernest}}, \bibinfo {author} {\bibfnamefont {S.}~\bibnamefont {Chakram}},
  \bibinfo {author} {\bibfnamefont {Y.}~\bibnamefont {Lu}}, \bibinfo {author}
  {\bibfnamefont {N.}~\bibnamefont {Irons}}, \bibinfo {author} {\bibfnamefont
  {R.~K.}\ \bibnamefont {Naik}}, \bibinfo {author} {\bibfnamefont
  {N.}~\bibnamefont {Leung}}, \bibinfo {author} {\bibfnamefont
  {J.}~\bibnamefont {Lawrence}}, \bibinfo {author} {\bibfnamefont
  {J.}~\bibnamefont {Koch}}, \ and\ \bibinfo {author} {\bibfnamefont {D.~I.}\
  \bibnamefont {Schuster}},\ }\href {https://arxiv.org/abs/1707.00656}
  {\bibfield  {journal} {\bibinfo  {journal} {arXiv:1707.00656 [quant-ph]}\ ,\
  } (\bibinfo {year} {2017})}\BibitemShut {NoStop}%
\bibitem [{\citenamefont {Gladchenko}\ \emph {et~al.}(2009)\citenamefont
  {Gladchenko}, \citenamefont {Olaya}, \citenamefont {Dupont-Ferrier},
  \citenamefont {Dou{\c c}ot}, \citenamefont {Ioffe},\ and\ \citenamefont
  {Gershenson}}]{gladchenko_superconducting_2009}%
  \BibitemOpen
  \bibfield  {author} {\bibinfo {author} {\bibfnamefont {S.}~\bibnamefont
  {Gladchenko}}, \bibinfo {author} {\bibfnamefont {D.}~\bibnamefont {Olaya}},
  \bibinfo {author} {\bibfnamefont {E.}~\bibnamefont {Dupont-Ferrier}},
  \bibinfo {author} {\bibfnamefont {B.}~\bibnamefont {Dou{\c c}ot}}, \bibinfo
  {author} {\bibfnamefont {L.~B.}\ \bibnamefont {Ioffe}}, \ and\ \bibinfo
  {author} {\bibfnamefont {M.~E.}\ \bibnamefont {Gershenson}},\ }\href
  {\doibase 10.1038/nphys1151} {\bibfield  {journal} {\bibinfo  {journal} {Nat.
  Phys.}\ }\textbf {\bibinfo {volume} {5}},\ \bibinfo {pages} {48} (\bibinfo
  {year} {2009})}\BibitemShut {NoStop}%
\bibitem [{\citenamefont {Brooks}\ \emph {et~al.}(2013)\citenamefont {Brooks},
  \citenamefont {Kitaev},\ and\ \citenamefont
  {Preskill}}]{brooks_protected_2013}%
  \BibitemOpen
  \bibfield  {author} {\bibinfo {author} {\bibfnamefont {P.}~\bibnamefont
  {Brooks}}, \bibinfo {author} {\bibfnamefont {A.}~\bibnamefont {Kitaev}}, \
  and\ \bibinfo {author} {\bibfnamefont {J.}~\bibnamefont {Preskill}},\ }\href
  {\doibase 10.1103/PhysRevA.87.052306} {\bibfield  {journal} {\bibinfo
  {journal} {Phys. Rev. A}\ }\textbf {\bibinfo {volume} {87}},\  (\bibinfo
  {year} {2013})}\BibitemShut {NoStop}%
\bibitem [{\citenamefont {Richer}\ \emph {et~al.}(2017)\citenamefont {Richer},
  \citenamefont {Maleeva}, \citenamefont {Skacel}, \citenamefont {Pop},\ and\
  \citenamefont {DiVincenzo}}]{richer_inductively_2017}%
  \BibitemOpen
  \bibfield  {author} {\bibinfo {author} {\bibfnamefont {S.}~\bibnamefont
  {Richer}}, \bibinfo {author} {\bibfnamefont {N.}~\bibnamefont {Maleeva}},
  \bibinfo {author} {\bibfnamefont {S.~T.}\ \bibnamefont {Skacel}}, \bibinfo
  {author} {\bibfnamefont {I.~M.}\ \bibnamefont {Pop}}, \ and\ \bibinfo
  {author} {\bibfnamefont {D.}~\bibnamefont {DiVincenzo}},\ }\href {\doibase
  10.1103/PhysRevB.96.174520} {\bibfield  {journal} {\bibinfo  {journal} {Phys.
  Rev. B}\ }\textbf {\bibinfo {volume} {96}},\  (\bibinfo {year}
  {2017})}\BibitemShut {NoStop}%
\bibitem [{\citenamefont {Groszkowski}\ \emph {et~al.}(2017)\citenamefont
  {Groszkowski}, \citenamefont {Di~Paolo}, \citenamefont {Grimsmo},
  \citenamefont {Blais}, \citenamefont {Schuster}, \citenamefont {Houck},\ and\
  \citenamefont {Koch}}]{groszkowski_coherence_2017}%
  \BibitemOpen
  \bibfield  {author} {\bibinfo {author} {\bibfnamefont {P.}~\bibnamefont
  {Groszkowski}}, \bibinfo {author} {\bibfnamefont {A.}~\bibnamefont
  {Di~Paolo}}, \bibinfo {author} {\bibfnamefont {A.~L.}\ \bibnamefont
  {Grimsmo}}, \bibinfo {author} {\bibfnamefont {A.}~\bibnamefont {Blais}},
  \bibinfo {author} {\bibfnamefont {D.~I.}\ \bibnamefont {Schuster}}, \bibinfo
  {author} {\bibfnamefont {A.~A.}\ \bibnamefont {Houck}}, \ and\ \bibinfo
  {author} {\bibfnamefont {J.}~\bibnamefont {Koch}},\ }\href
  {https://arxiv.org/abs/1708.02886} {\bibfield  {journal} {\bibinfo  {journal}
  {arXiv:1708.02886 [quant-ph]}\ ,\ } (\bibinfo {year} {2017})}\BibitemShut
  {NoStop}%
\bibitem [{\citenamefont {Petrescu}\ \emph {et~al.}(2017)\citenamefont
  {Petrescu}, \citenamefont {T{\"u}reci}, \citenamefont {Ustinov},\ and\
  \citenamefont {Pop}}]{petrescu_fluxon-based_2017}%
  \BibitemOpen
  \bibfield  {author} {\bibinfo {author} {\bibfnamefont {A.}~\bibnamefont
  {Petrescu}}, \bibinfo {author} {\bibfnamefont {H.~E.}\ \bibnamefont
  {T{\"u}reci}}, \bibinfo {author} {\bibfnamefont {A.~V.}\ \bibnamefont
  {Ustinov}}, \ and\ \bibinfo {author} {\bibfnamefont {I.~M.}\ \bibnamefont
  {Pop}},\ }\href {https://arxiv.org/abs/1712.08630} {\bibfield  {journal}
  {\bibinfo  {journal} {arXiv:1712.08630 [cond-mat.mes-hall]}\ ,\ } (\bibinfo
  {year} {2017})}\BibitemShut {NoStop}%
\bibitem [{\citenamefont {Astafiev}\ \emph {et~al.}(2012)\citenamefont
  {Astafiev}, \citenamefont {Ioffe}, \citenamefont {Kafanov}, \citenamefont
  {Pashkin}, \citenamefont {Arutyunov}, \citenamefont {Shahar}, \citenamefont
  {Cohen},\ and\ \citenamefont {Tsai}}]{astafiev_coherent_2012}%
  \BibitemOpen
  \bibfield  {author} {\bibinfo {author} {\bibfnamefont {O.~V.}\ \bibnamefont
  {Astafiev}}, \bibinfo {author} {\bibfnamefont {L.~B.}\ \bibnamefont {Ioffe}},
  \bibinfo {author} {\bibfnamefont {S.}~\bibnamefont {Kafanov}}, \bibinfo
  {author} {\bibfnamefont {Y.~A.}\ \bibnamefont {Pashkin}}, \bibinfo {author}
  {\bibfnamefont {K.~Y.}\ \bibnamefont {Arutyunov}}, \bibinfo {author}
  {\bibfnamefont {D.}~\bibnamefont {Shahar}}, \bibinfo {author} {\bibfnamefont
  {O.}~\bibnamefont {Cohen}}, \ and\ \bibinfo {author} {\bibfnamefont {J.~S.}\
  \bibnamefont {Tsai}},\ }\href {\doibase 10.1038/nature10930} {\bibfield
  {journal} {\bibinfo  {journal} {Nature}\ }\textbf {\bibinfo {volume} {484}},\
  \bibinfo {pages} {355} (\bibinfo {year} {2012})}\BibitemShut {NoStop}%
\bibitem [{\citenamefont {Belkin}\ \emph {et~al.}(2015)\citenamefont {Belkin},
  \citenamefont {Belkin}, \citenamefont {Vakaryuk}, \citenamefont
  {Khlebnikov},\ and\ \citenamefont {Bezryadin}}]{belkin_formation_2015}%
  \BibitemOpen
  \bibfield  {author} {\bibinfo {author} {\bibfnamefont {A.}~\bibnamefont
  {Belkin}}, \bibinfo {author} {\bibfnamefont {M.}~\bibnamefont {Belkin}},
  \bibinfo {author} {\bibfnamefont {V.}~\bibnamefont {Vakaryuk}}, \bibinfo
  {author} {\bibfnamefont {S.}~\bibnamefont {Khlebnikov}}, \ and\ \bibinfo
  {author} {\bibfnamefont {A.}~\bibnamefont {Bezryadin}},\ }\href {\doibase
  10.1103/PhysRevX.5.021023} {\bibfield  {journal} {\bibinfo  {journal} {Phys.
  Rev. X}\ }\textbf {\bibinfo {volume} {5}},\  (\bibinfo {year}
  {2015})}\BibitemShut {NoStop}%
\bibitem [{\citenamefont {Bell}\ \emph {et~al.}(2016)\citenamefont {Bell},
  \citenamefont {Zhang}, \citenamefont {Ioffe},\ and\ \citenamefont
  {Gershenson}}]{bell_spectroscopic_2016}%
  \BibitemOpen
  \bibfield  {author} {\bibinfo {author} {\bibfnamefont {M.~T.}\ \bibnamefont
  {Bell}}, \bibinfo {author} {\bibfnamefont {W.}~\bibnamefont {Zhang}},
  \bibinfo {author} {\bibfnamefont {L.~B.}\ \bibnamefont {Ioffe}}, \ and\
  \bibinfo {author} {\bibfnamefont {M.~E.}\ \bibnamefont {Gershenson}},\ }\href
  {\doibase 10.1103/PhysRevLett.116.107002} {\bibfield  {journal} {\bibinfo
  {journal} {Phys. Rev. Lett.}\ }\textbf {\bibinfo {volume} {116}},\  (\bibinfo
  {year} {2016})}\BibitemShut {NoStop}%
\bibitem [{\citenamefont {M{\"u}ller}\ \emph {et~al.}(2017)\citenamefont
  {M{\"u}ller}, \citenamefont {Guan}, \citenamefont {Vogt}, \citenamefont
  {Cole},\ and\ \citenamefont {Stace}}]{muller_passive_2017}%
  \BibitemOpen
  \bibfield  {author} {\bibinfo {author} {\bibfnamefont {C.}~\bibnamefont
  {M{\"u}ller}}, \bibinfo {author} {\bibfnamefont {S.}~\bibnamefont {Guan}},
  \bibinfo {author} {\bibfnamefont {N.}~\bibnamefont {Vogt}}, \bibinfo {author}
  {\bibfnamefont {J.~H.}\ \bibnamefont {Cole}}, \ and\ \bibinfo {author}
  {\bibfnamefont {T.~M.}\ \bibnamefont {Stace}},\ }\href
  {https://arxiv.org/abs/1709.09826} {\bibfield  {journal} {\bibinfo  {journal}
  {arXiv:1709.09826 [quant-ph]}\ ,\ } (\bibinfo {year} {2017})}\BibitemShut
  {NoStop}%
\bibitem [{\citenamefont {Ho~Eom}\ \emph {et~al.}(2012)\citenamefont {Ho~Eom},
  \citenamefont {Day}, \citenamefont {LeDuc},\ and\ \citenamefont
  {Zmuidzinas}}]{ho_eom_wideband_2012}%
  \BibitemOpen
  \bibfield  {author} {\bibinfo {author} {\bibfnamefont {B.}~\bibnamefont
  {Ho~Eom}}, \bibinfo {author} {\bibfnamefont {P.~K.}\ \bibnamefont {Day}},
  \bibinfo {author} {\bibfnamefont {H.~G.}\ \bibnamefont {LeDuc}}, \ and\
  \bibinfo {author} {\bibfnamefont {J.}~\bibnamefont {Zmuidzinas}},\ }\href
  {\doibase 10.1038/nphys2356} {\bibfield  {journal} {\bibinfo  {journal} {Nat.
  Phys.}\ }\textbf {\bibinfo {volume} {8}},\ \bibinfo {pages} {623} (\bibinfo
  {year} {2012})}\BibitemShut {NoStop}%
\bibitem [{\citenamefont {Vissers}\ \emph {et~al.}(2016)\citenamefont
  {Vissers}, \citenamefont {Erickson}, \citenamefont {Ku}, \citenamefont
  {Vale}, \citenamefont {Wu}, \citenamefont {Hilton},\ and\ \citenamefont
  {Pappas}}]{vissers_low-noise_2016}%
  \BibitemOpen
  \bibfield  {author} {\bibinfo {author} {\bibfnamefont {M.~R.}\ \bibnamefont
  {Vissers}}, \bibinfo {author} {\bibfnamefont {R.~P.}\ \bibnamefont
  {Erickson}}, \bibinfo {author} {\bibfnamefont {H.-S.}\ \bibnamefont {Ku}},
  \bibinfo {author} {\bibfnamefont {L.}~\bibnamefont {Vale}}, \bibinfo {author}
  {\bibfnamefont {X.}~\bibnamefont {Wu}}, \bibinfo {author} {\bibfnamefont
  {G.~C.}\ \bibnamefont {Hilton}}, \ and\ \bibinfo {author} {\bibfnamefont
  {D.~P.}\ \bibnamefont {Pappas}},\ }\href {\doibase 10.1063/1.4937922}
  {\bibfield  {journal} {\bibinfo  {journal} {Appl. Phys. Lett.}\ }\textbf
  {\bibinfo {volume} {108}},\ \bibinfo {pages} {012601} (\bibinfo {year}
  {2016})}\BibitemShut {NoStop}%
\bibitem [{\citenamefont {Cohen}\ \emph {et~al.}(2017)\citenamefont {Cohen},
  \citenamefont {Smith}, \citenamefont {Devoret},\ and\ \citenamefont
  {Mirrahimi}}]{cohen_degeneracy-preserving_2017}%
  \BibitemOpen
  \bibfield  {author} {\bibinfo {author} {\bibfnamefont {J.}~\bibnamefont
  {Cohen}}, \bibinfo {author} {\bibfnamefont {W.~C.}\ \bibnamefont {Smith}},
  \bibinfo {author} {\bibfnamefont {M.~H.}\ \bibnamefont {Devoret}}, \ and\
  \bibinfo {author} {\bibfnamefont {M.}~\bibnamefont {Mirrahimi}},\ }\href
  {\doibase 10.1103/PhysRevLett.119.060503} {\bibfield  {journal} {\bibinfo
  {journal} {Phys. Rev. Lett.}\ }\textbf {\bibinfo {volume} {119}},\  (\bibinfo
  {year} {2017})}\BibitemShut {NoStop}%
\bibitem [{\citenamefont {Puri}\ \emph {et~al.}(2017)\citenamefont {Puri},
  \citenamefont {Boutin},\ and\ \citenamefont {Blais}}]{puri_engineering_2017}%
  \BibitemOpen
  \bibfield  {author} {\bibinfo {author} {\bibfnamefont {S.}~\bibnamefont
  {Puri}}, \bibinfo {author} {\bibfnamefont {S.}~\bibnamefont {Boutin}}, \ and\
  \bibinfo {author} {\bibfnamefont {A.}~\bibnamefont {Blais}},\ }\href
  {\doibase 10.1038/s41534-017-0019-1} {\bibfield  {journal} {\bibinfo
  {journal} {npj Quantum Inf.}\ }\textbf {\bibinfo {volume} {3}},\  (\bibinfo
  {year} {2017})}\BibitemShut {NoStop}%
\bibitem [{\citenamefont {Day}\ \emph {et~al.}(2003)\citenamefont {Day},
  \citenamefont {LeDuc}, \citenamefont {Mazin}, \citenamefont {Vayonakis},\
  and\ \citenamefont {Zmuidzinas}}]{day_broadband_2003}%
  \BibitemOpen
  \bibfield  {author} {\bibinfo {author} {\bibfnamefont {P.~K.}\ \bibnamefont
  {Day}}, \bibinfo {author} {\bibfnamefont {H.~G.}\ \bibnamefont {LeDuc}},
  \bibinfo {author} {\bibfnamefont {B.~A.}\ \bibnamefont {Mazin}}, \bibinfo
  {author} {\bibfnamefont {A.}~\bibnamefont {Vayonakis}}, \ and\ \bibinfo
  {author} {\bibfnamefont {J.}~\bibnamefont {Zmuidzinas}},\ }\href {\doibase
  10.1038/nature02037} {\bibfield  {journal} {\bibinfo  {journal} {Nature}\
  }\textbf {\bibinfo {volume} {425}},\ \bibinfo {pages} {817} (\bibinfo {year}
  {2003})}\BibitemShut {NoStop}%
\bibitem [{\citenamefont {Aumentado}\ \emph {et~al.}(2004)\citenamefont
  {Aumentado}, \citenamefont {Keller}, \citenamefont {Martinis},\ and\
  \citenamefont {Devoret}}]{aumentado_nonequilibrium_2004}%
  \BibitemOpen
  \bibfield  {author} {\bibinfo {author} {\bibfnamefont {J.}~\bibnamefont
  {Aumentado}}, \bibinfo {author} {\bibfnamefont {M.~W.}\ \bibnamefont
  {Keller}}, \bibinfo {author} {\bibfnamefont {J.~M.}\ \bibnamefont
  {Martinis}}, \ and\ \bibinfo {author} {\bibfnamefont {M.~H.}\ \bibnamefont
  {Devoret}},\ }\href {\doibase 10.1103/PhysRevLett.92.066802} {\bibfield
  {journal} {\bibinfo  {journal} {Phys. Rev. Lett.}\ }\textbf {\bibinfo
  {volume} {92}},\  (\bibinfo {year} {2004})}\BibitemShut {NoStop}%
\bibitem [{\citenamefont {{de Visser}}\ \emph {et~al.}(2011)\citenamefont {{de
  Visser}}, \citenamefont {Baselmans}, \citenamefont {Diener}, \citenamefont
  {Yates}, \citenamefont {Endo},\ and\ \citenamefont
  {Klapwijk}}]{de_visser_number_2011}%
  \BibitemOpen
  \bibfield  {author} {\bibinfo {author} {\bibfnamefont {P.~J.}\ \bibnamefont
  {{de Visser}}}, \bibinfo {author} {\bibfnamefont {J.~J.~A.}\ \bibnamefont
  {Baselmans}}, \bibinfo {author} {\bibfnamefont {P.}~\bibnamefont {Diener}},
  \bibinfo {author} {\bibfnamefont {S.~J.~C.}\ \bibnamefont {Yates}}, \bibinfo
  {author} {\bibfnamefont {A.}~\bibnamefont {Endo}}, \ and\ \bibinfo {author}
  {\bibfnamefont {T.~M.}\ \bibnamefont {Klapwijk}},\ }\href {\doibase
  10.1103/PhysRevLett.106.167004} {\bibfield  {journal} {\bibinfo  {journal}
  {Phys. Rev. Lett.}\ }\textbf {\bibinfo {volume} {106}},\  (\bibinfo {year}
  {2011})}\BibitemShut {NoStop}%
\bibitem [{\citenamefont {Maisi}\ \emph {et~al.}(2013)\citenamefont {Maisi},
  \citenamefont {Lotkhov}, \citenamefont {Kemppinen}, \citenamefont {Heimes},
  \citenamefont {Muhonen},\ and\ \citenamefont
  {Pekola}}]{maisi_excitation_2013}%
  \BibitemOpen
  \bibfield  {author} {\bibinfo {author} {\bibfnamefont {V.~F.}\ \bibnamefont
  {Maisi}}, \bibinfo {author} {\bibfnamefont {S.~V.}\ \bibnamefont {Lotkhov}},
  \bibinfo {author} {\bibfnamefont {A.}~\bibnamefont {Kemppinen}}, \bibinfo
  {author} {\bibfnamefont {A.}~\bibnamefont {Heimes}}, \bibinfo {author}
  {\bibfnamefont {J.~T.}\ \bibnamefont {Muhonen}}, \ and\ \bibinfo {author}
  {\bibfnamefont {J.~P.}\ \bibnamefont {Pekola}},\ }\href {\doibase
  10.1103/PhysRevLett.111.147001} {\bibfield  {journal} {\bibinfo  {journal}
  {Phys. Rev. Lett.}\ }\textbf {\bibinfo {volume} {111}},\  (\bibinfo {year}
  {2013})}\BibitemShut {NoStop}%
\bibitem [{\citenamefont {Levenson-Falk}\ \emph {et~al.}(2014)\citenamefont
  {Levenson-Falk}, \citenamefont {Kos}, \citenamefont {Vijay}, \citenamefont
  {Glazman},\ and\ \citenamefont
  {Siddiqi}}]{levenson-falk_single-quasiparticle_2014}%
  \BibitemOpen
  \bibfield  {author} {\bibinfo {author} {\bibfnamefont {E.~M.}\ \bibnamefont
  {Levenson-Falk}}, \bibinfo {author} {\bibfnamefont {F.}~\bibnamefont {Kos}},
  \bibinfo {author} {\bibfnamefont {R.}~\bibnamefont {Vijay}}, \bibinfo
  {author} {\bibfnamefont {L.}~\bibnamefont {Glazman}}, \ and\ \bibinfo
  {author} {\bibfnamefont {I.}~\bibnamefont {Siddiqi}},\ }\href {\doibase
  10.1103/PhysRevLett.112.047002} {\bibfield  {journal} {\bibinfo  {journal}
  {Phys. Rev. Lett.}\ }\textbf {\bibinfo {volume} {112}},\  (\bibinfo {year}
  {2014})}\BibitemShut {NoStop}%
\bibitem [{\citenamefont {Wang}\ \emph {et~al.}(2014)\citenamefont {Wang},
  \citenamefont {Gao}, \citenamefont {Pop}, \citenamefont {Vool}, \citenamefont
  {Axline}, \citenamefont {Brecht}, \citenamefont {Heeres}, \citenamefont
  {Frunzio}, \citenamefont {Devoret}, \citenamefont {Catelani}, \citenamefont
  {Glazman},\ and\ \citenamefont {Schoelkopf}}]{Wang_NatComm2014}%
  \BibitemOpen
  \bibfield  {author} {\bibinfo {author} {\bibfnamefont {C.}~\bibnamefont
  {Wang}}, \bibinfo {author} {\bibfnamefont {Y.~Y.}\ \bibnamefont {Gao}},
  \bibinfo {author} {\bibfnamefont {I.~M.}\ \bibnamefont {Pop}}, \bibinfo
  {author} {\bibfnamefont {U.}~\bibnamefont {Vool}}, \bibinfo {author}
  {\bibfnamefont {C.}~\bibnamefont {Axline}}, \bibinfo {author} {\bibfnamefont
  {T.}~\bibnamefont {Brecht}}, \bibinfo {author} {\bibfnamefont {R.~W.}\
  \bibnamefont {Heeres}}, \bibinfo {author} {\bibfnamefont {L.}~\bibnamefont
  {Frunzio}}, \bibinfo {author} {\bibfnamefont {M.~H.}\ \bibnamefont
  {Devoret}}, \bibinfo {author} {\bibfnamefont {G.}~\bibnamefont {Catelani}},
  \bibinfo {author} {\bibfnamefont {L.~I.}\ \bibnamefont {Glazman}}, \ and\
  \bibinfo {author} {\bibfnamefont {R.~J.}\ \bibnamefont {Schoelkopf}},\ }\href
  {\doibase 10.1038/ncomms6836} {\bibfield  {journal} {\bibinfo  {journal}
  {Nat. Commun.}\ }\textbf {\bibinfo {volume} {5}},\ \bibinfo {pages} {5836}
  (\bibinfo {year} {2014})}\BibitemShut {NoStop}%
\bibitem [{\citenamefont {Bilmes}\ \emph {et~al.}(2017)\citenamefont {Bilmes},
  \citenamefont {Zanker}, \citenamefont {Heimes}, \citenamefont {Marthaler},
  \citenamefont {Sch{\"o}n}, \citenamefont {Weiss}, \citenamefont {Ustinov},\
  and\ \citenamefont {Lisenfeld}}]{bilmes_electronic_2017}%
  \BibitemOpen
  \bibfield  {author} {\bibinfo {author} {\bibfnamefont {A.}~\bibnamefont
  {Bilmes}}, \bibinfo {author} {\bibfnamefont {S.}~\bibnamefont {Zanker}},
  \bibinfo {author} {\bibfnamefont {A.}~\bibnamefont {Heimes}}, \bibinfo
  {author} {\bibfnamefont {M.}~\bibnamefont {Marthaler}}, \bibinfo {author}
  {\bibfnamefont {G.}~\bibnamefont {Sch{\"o}n}}, \bibinfo {author}
  {\bibfnamefont {G.}~\bibnamefont {Weiss}}, \bibinfo {author} {\bibfnamefont
  {A.~V.}\ \bibnamefont {Ustinov}}, \ and\ \bibinfo {author} {\bibfnamefont
  {J.}~\bibnamefont {Lisenfeld}},\ }\href {\doibase 10.1103/PhysRevB.96.064504}
  {\bibfield  {journal} {\bibinfo  {journal} {Phys. Rev. B}\ }\textbf {\bibinfo
  {volume} {96}},\  (\bibinfo {year} {2017})}\BibitemShut {NoStop}%
\bibitem [{\citenamefont {Nsanzineza}\ and\ \citenamefont
  {Plourde}(2014)}]{nsanzineza_trapping_2014}%
  \BibitemOpen
  \bibfield  {author} {\bibinfo {author} {\bibfnamefont {I.}~\bibnamefont
  {Nsanzineza}}\ and\ \bibinfo {author} {\bibfnamefont {B.~L.~T.}\ \bibnamefont
  {Plourde}},\ }\href {\doibase 10.1103/PhysRevLett.113.117002} {\bibfield
  {journal} {\bibinfo  {journal} {Phys. Rev. Lett.}\ }\textbf {\bibinfo
  {volume} {113}},\  (\bibinfo {year} {2014})}\BibitemShut {NoStop}%
\bibitem [{\citenamefont {Janvier}\ \emph {et~al.}(2015)\citenamefont
  {Janvier}, \citenamefont {Tosi}, \citenamefont {Bretheau}, \citenamefont
  {{\"o}.~Girit}, \citenamefont {Stern}, \citenamefont {Bertet}, \citenamefont
  {Joyez}, \citenamefont {Vion}, \citenamefont {Esteve}, \citenamefont
  {Goffman}, \citenamefont {Pothier},\ and\ \citenamefont
  {Urbina}}]{janvier_coherent_2015}%
  \BibitemOpen
  \bibfield  {author} {\bibinfo {author} {\bibfnamefont {C.}~\bibnamefont
  {Janvier}}, \bibinfo {author} {\bibfnamefont {L.}~\bibnamefont {Tosi}},
  \bibinfo {author} {\bibfnamefont {L.}~\bibnamefont {Bretheau}}, \bibinfo
  {author} {\bibfnamefont {{\c c}.}~\bibnamefont {{\"o}.~Girit}}, \bibinfo
  {author} {\bibfnamefont {M.}~\bibnamefont {Stern}}, \bibinfo {author}
  {\bibfnamefont {P.}~\bibnamefont {Bertet}}, \bibinfo {author} {\bibfnamefont
  {P.}~\bibnamefont {Joyez}}, \bibinfo {author} {\bibfnamefont
  {D.}~\bibnamefont {Vion}}, \bibinfo {author} {\bibfnamefont {D.}~\bibnamefont
  {Esteve}}, \bibinfo {author} {\bibfnamefont {M.~F.}\ \bibnamefont {Goffman}},
  \bibinfo {author} {\bibfnamefont {H.}~\bibnamefont {Pothier}}, \ and\
  \bibinfo {author} {\bibfnamefont {C.}~\bibnamefont {Urbina}},\ }\href
  {\doibase 10.1126/science.aab2179} {\bibfield  {journal} {\bibinfo  {journal}
  {Science}\ }\textbf {\bibinfo {volume} {349}},\ \bibinfo {pages} {1199}
  (\bibinfo {year} {2015})}\BibitemShut {NoStop}%
\bibitem [{\citenamefont {Gustavsson}\ \emph {et~al.}(2016)\citenamefont
  {Gustavsson}, \citenamefont {Yan}, \citenamefont {Catelani}, \citenamefont
  {Bylander}, \citenamefont {Kamal}, \citenamefont {Birenbaum}, \citenamefont
  {Hover}, \citenamefont {Rosenberg}, \citenamefont {Samach}, \citenamefont
  {Sears}, \citenamefont {Weber}, \citenamefont {Yoder}, \citenamefont
  {Clarke}, \citenamefont {Kerman}, \citenamefont {Yoshihara}, \citenamefont
  {Nakamura}, \citenamefont {Orlando},\ and\ \citenamefont
  {Oliver}}]{gustavsson_suppressing_2016}%
  \BibitemOpen
  \bibfield  {author} {\bibinfo {author} {\bibfnamefont {S.}~\bibnamefont
  {Gustavsson}}, \bibinfo {author} {\bibfnamefont {F.}~\bibnamefont {Yan}},
  \bibinfo {author} {\bibfnamefont {G.}~\bibnamefont {Catelani}}, \bibinfo
  {author} {\bibfnamefont {J.}~\bibnamefont {Bylander}}, \bibinfo {author}
  {\bibfnamefont {A.}~\bibnamefont {Kamal}}, \bibinfo {author} {\bibfnamefont
  {J.}~\bibnamefont {Birenbaum}}, \bibinfo {author} {\bibfnamefont
  {D.}~\bibnamefont {Hover}}, \bibinfo {author} {\bibfnamefont
  {D.}~\bibnamefont {Rosenberg}}, \bibinfo {author} {\bibfnamefont
  {G.}~\bibnamefont {Samach}}, \bibinfo {author} {\bibfnamefont {A.~P.}\
  \bibnamefont {Sears}}, \bibinfo {author} {\bibfnamefont {S.~J.}\ \bibnamefont
  {Weber}}, \bibinfo {author} {\bibfnamefont {J.~L.}\ \bibnamefont {Yoder}},
  \bibinfo {author} {\bibfnamefont {J.}~\bibnamefont {Clarke}}, \bibinfo
  {author} {\bibfnamefont {A.~J.}\ \bibnamefont {Kerman}}, \bibinfo {author}
  {\bibfnamefont {F.}~\bibnamefont {Yoshihara}}, \bibinfo {author}
  {\bibfnamefont {Y.}~\bibnamefont {Nakamura}}, \bibinfo {author}
  {\bibfnamefont {T.~P.}\ \bibnamefont {Orlando}}, \ and\ \bibinfo {author}
  {\bibfnamefont {W.~D.}\ \bibnamefont {Oliver}},\ }\href {\doibase
  10.1126/science.aah5844} {\bibfield  {journal} {\bibinfo  {journal}
  {Science}\ }\textbf {\bibinfo {volume} {354}},\ \bibinfo {pages} {1573}
  (\bibinfo {year} {2016})}\BibitemShut {NoStop}%
\bibitem [{\citenamefont {Taupin}\ \emph {et~al.}(2016)\citenamefont {Taupin},
  \citenamefont {Khaymovich}, \citenamefont {Meschke}, \citenamefont
  {Mel'nikov},\ and\ \citenamefont {Pekola}}]{taupin_tunable_2016}%
  \BibitemOpen
  \bibfield  {author} {\bibinfo {author} {\bibfnamefont {M.}~\bibnamefont
  {Taupin}}, \bibinfo {author} {\bibfnamefont {I.~M.}\ \bibnamefont
  {Khaymovich}}, \bibinfo {author} {\bibfnamefont {M.}~\bibnamefont {Meschke}},
  \bibinfo {author} {\bibfnamefont {A.~S.}\ \bibnamefont {Mel'nikov}}, \ and\
  \bibinfo {author} {\bibfnamefont {J.~P.}\ \bibnamefont {Pekola}},\ }\href
  {\doibase 10.1038/ncomms10977} {\bibfield  {journal} {\bibinfo  {journal}
  {Nat. Commun.}\ }\textbf {\bibinfo {volume} {7}},\ \bibinfo {pages} {10977}
  (\bibinfo {year} {2016})}\BibitemShut {NoStop}%
\bibitem [{\citenamefont {Higginbotham}\ \emph {et~al.}(2015)\citenamefont
  {Higginbotham}, \citenamefont {Albrecht}, \citenamefont {Kir{\v s}anskas},
  \citenamefont {Chang}, \citenamefont {Kuemmeth}, \citenamefont {Krogstrup},
  \citenamefont {Jespersen}, \citenamefont {Nyg{\aa}rd}, \citenamefont
  {Flensberg},\ and\ \citenamefont {Marcus}}]{higginbotham_parity_2015}%
  \BibitemOpen
  \bibfield  {author} {\bibinfo {author} {\bibfnamefont {A.~P.}\ \bibnamefont
  {Higginbotham}}, \bibinfo {author} {\bibfnamefont {S.~M.}\ \bibnamefont
  {Albrecht}}, \bibinfo {author} {\bibfnamefont {G.}~\bibnamefont {Kir{\v
  s}anskas}}, \bibinfo {author} {\bibfnamefont {W.}~\bibnamefont {Chang}},
  \bibinfo {author} {\bibfnamefont {F.}~\bibnamefont {Kuemmeth}}, \bibinfo
  {author} {\bibfnamefont {P.}~\bibnamefont {Krogstrup}}, \bibinfo {author}
  {\bibfnamefont {T.~S.}\ \bibnamefont {Jespersen}}, \bibinfo {author}
  {\bibfnamefont {J.}~\bibnamefont {Nyg{\aa}rd}}, \bibinfo {author}
  {\bibfnamefont {K.}~\bibnamefont {Flensberg}}, \ and\ \bibinfo {author}
  {\bibfnamefont {C.~M.}\ \bibnamefont {Marcus}},\ }\href {\doibase
  10.1038/nphys3461} {\bibfield  {journal} {\bibinfo  {journal} {Nat. Phys.}\
  }\textbf {\bibinfo {volume} {11}},\ \bibinfo {pages} {1017} (\bibinfo {year}
  {2015})}\BibitemShut {NoStop}%
\bibitem [{\citenamefont {Aasen}\ \emph {et~al.}(2016)\citenamefont {Aasen},
  \citenamefont {Hell}, \citenamefont {Mishmash}, \citenamefont {Higginbotham},
  \citenamefont {Danon}, \citenamefont {Leijnse}, \citenamefont {Jespersen},
  \citenamefont {Folk}, \citenamefont {Marcus}, \citenamefont {Flensberg},\
  and\ \citenamefont {Alicea}}]{aasen_milestones_2016}%
  \BibitemOpen
  \bibfield  {author} {\bibinfo {author} {\bibfnamefont {D.}~\bibnamefont
  {Aasen}}, \bibinfo {author} {\bibfnamefont {M.}~\bibnamefont {Hell}},
  \bibinfo {author} {\bibfnamefont {R.~V.}\ \bibnamefont {Mishmash}}, \bibinfo
  {author} {\bibfnamefont {A.}~\bibnamefont {Higginbotham}}, \bibinfo {author}
  {\bibfnamefont {J.}~\bibnamefont {Danon}}, \bibinfo {author} {\bibfnamefont
  {M.}~\bibnamefont {Leijnse}}, \bibinfo {author} {\bibfnamefont {T.~S.}\
  \bibnamefont {Jespersen}}, \bibinfo {author} {\bibfnamefont {J.~A.}\
  \bibnamefont {Folk}}, \bibinfo {author} {\bibfnamefont {C.~M.}\ \bibnamefont
  {Marcus}}, \bibinfo {author} {\bibfnamefont {K.}~\bibnamefont {Flensberg}}, \
  and\ \bibinfo {author} {\bibfnamefont {J.}~\bibnamefont {Alicea}},\ }\href
  {\doibase 10.1103/PhysRevX.6.031016} {\bibfield  {journal} {\bibinfo
  {journal} {Phys. Rev. X}\ }\textbf {\bibinfo {volume} {6}},\  (\bibinfo
  {year} {2016})}\BibitemShut {NoStop}%
\bibitem [{\citenamefont {Zoepfl}\ \emph {et~al.}(2017)\citenamefont {Zoepfl},
  \citenamefont {Muppalla}, \citenamefont {Schneider}, \citenamefont
  {Kasemann}, \citenamefont {Partel},\ and\ \citenamefont
  {Kirchmair}}]{zoepfl_characterization_2017}%
  \BibitemOpen
  \bibfield  {author} {\bibinfo {author} {\bibfnamefont {D.}~\bibnamefont
  {Zoepfl}}, \bibinfo {author} {\bibfnamefont {P.~R.}\ \bibnamefont
  {Muppalla}}, \bibinfo {author} {\bibfnamefont {C.~M.~F.}\ \bibnamefont
  {Schneider}}, \bibinfo {author} {\bibfnamefont {S.}~\bibnamefont {Kasemann}},
  \bibinfo {author} {\bibfnamefont {S.}~\bibnamefont {Partel}}, \ and\ \bibinfo
  {author} {\bibfnamefont {G.}~\bibnamefont {Kirchmair}},\ }\href {\doibase
  10.1063/1.4992070} {\bibfield  {journal} {\bibinfo  {journal} {AIP Adv.}\
  }\textbf {\bibinfo {volume} {7}},\ \bibinfo {pages} {085118} (\bibinfo {year}
  {2017})}\BibitemShut {NoStop}%
\bibitem [{\citenamefont {Kou}\ \emph {et~al.}(2017)\citenamefont {Kou},
  \citenamefont {Smith}, \citenamefont {Vool}, \citenamefont {Pop},
  \citenamefont {Sliwa}, \citenamefont {Hatridge}, \citenamefont {Frunzio},\
  and\ \citenamefont {Devoret}}]{kou_simultaneous_2017}%
  \BibitemOpen
  \bibfield  {author} {\bibinfo {author} {\bibfnamefont {A.}~\bibnamefont
  {Kou}}, \bibinfo {author} {\bibfnamefont {W.~C.}\ \bibnamefont {Smith}},
  \bibinfo {author} {\bibfnamefont {U.}~\bibnamefont {Vool}}, \bibinfo {author}
  {\bibfnamefont {I.~M.}\ \bibnamefont {Pop}}, \bibinfo {author} {\bibfnamefont
  {K.~M.}\ \bibnamefont {Sliwa}}, \bibinfo {author} {\bibfnamefont {M.~H.}\
  \bibnamefont {Hatridge}}, \bibinfo {author} {\bibfnamefont {L.}~\bibnamefont
  {Frunzio}}, \ and\ \bibinfo {author} {\bibfnamefont {M.~H.}\ \bibnamefont
  {Devoret}},\ }\href {https://arxiv.org/abs/1705.05712} {\bibfield  {journal}
  {\bibinfo  {journal} {arXiv:1705.05712 [quant-ph]}\ ,\ } (\bibinfo {year}
  {2017})}\BibitemShut {NoStop}%
\bibitem [{\citenamefont {Gao}(2008)}]{gao_physics_2008}%
  \BibitemOpen
  \bibfield  {author} {\bibinfo {author} {\bibfnamefont {J.}~\bibnamefont
  {Gao}},\ }\emph {\bibinfo {title} {The Physics of Superconducting Microwave
  Resonators}},\ \href
  {http://resolver.caltech.edu/CaltechETD:etd-06092008-235549} {\bibinfo {type}
  {Phd dissertation}},\ \bibinfo  {school} {California Institute of Technology}
  (\bibinfo {year} {2008})\BibitemShut {NoStop}%
\bibitem [{\citenamefont {Turneaure}\ \emph {et~al.}(1991)\citenamefont
  {Turneaure}, \citenamefont {Halbritter},\ and\ \citenamefont
  {Schwettman}}]{turneaure_surface_1991}%
  \BibitemOpen
  \bibfield  {author} {\bibinfo {author} {\bibfnamefont {J.~P.}\ \bibnamefont
  {Turneaure}}, \bibinfo {author} {\bibfnamefont {J.}~\bibnamefont
  {Halbritter}}, \ and\ \bibinfo {author} {\bibfnamefont {H.~A.}\ \bibnamefont
  {Schwettman}},\ }\href {\doibase 10.1007/BF00618215} {\bibfield  {journal}
  {\bibinfo  {journal} {J. Supercond.}\ }\textbf {\bibinfo {volume} {4}},\
  \bibinfo {pages} {341} (\bibinfo {year} {1991})}\BibitemShut {NoStop}%
\bibitem [{\citenamefont {Pracht}\ \emph {et~al.}(2016)\citenamefont {Pracht},
  \citenamefont {Bachar}, \citenamefont {Benfatto}, \citenamefont {Deutscher},
  \citenamefont {Farber}, \citenamefont {Dressel},\ and\ \citenamefont
  {Scheffler}}]{pracht_enhanced_2016}%
  \BibitemOpen
  \bibfield  {author} {\bibinfo {author} {\bibfnamefont {U.~S.}\ \bibnamefont
  {Pracht}}, \bibinfo {author} {\bibfnamefont {N.}~\bibnamefont {Bachar}},
  \bibinfo {author} {\bibfnamefont {L.}~\bibnamefont {Benfatto}}, \bibinfo
  {author} {\bibfnamefont {G.}~\bibnamefont {Deutscher}}, \bibinfo {author}
  {\bibfnamefont {E.}~\bibnamefont {Farber}}, \bibinfo {author} {\bibfnamefont
  {M.}~\bibnamefont {Dressel}}, \ and\ \bibinfo {author} {\bibfnamefont
  {M.}~\bibnamefont {Scheffler}},\ }\href {\doibase 10.1103/PhysRevB.93.100503}
  {\bibfield  {journal} {\bibinfo  {journal} {Phys. Rev. B}\ }\textbf {\bibinfo
  {volume} {93}},\  (\bibinfo {year} {2016})}\BibitemShut {NoStop}%
\bibitem [{\citenamefont {Dupr{\'e}}\ \emph {et~al.}(2018)\citenamefont
  {Dupr{\'e}}, \citenamefont {Beno{\^\i}t}, \citenamefont {Bideaud},
  \citenamefont {Bourrion}, \citenamefont {Calvo}, \citenamefont {Catalano},
  \citenamefont {Gomez}, \citenamefont {Goupy}, \citenamefont {Grenet},
  \citenamefont {Gr{\"u}nhaupt}, \citenamefont {Klein}, \citenamefont {{von
  L{\"u}pke}}, \citenamefont {Maleeva}, \citenamefont {Valenti}, \citenamefont
  {Monfardini}, \citenamefont {Pop},\ and\ \citenamefont
  {Levy-Bertrand}}]{dupre_phase_2017}%
  \BibitemOpen
  \bibfield  {author} {\bibinfo {author} {\bibfnamefont {O.}~\bibnamefont
  {Dupr{\'e}}}, \bibinfo {author} {\bibfnamefont {A.}~\bibnamefont
  {Beno{\^\i}t}}, \bibinfo {author} {\bibfnamefont {A.}~\bibnamefont
  {Bideaud}}, \bibinfo {author} {\bibfnamefont {O.}~\bibnamefont {Bourrion}},
  \bibinfo {author} {\bibfnamefont {M.}~\bibnamefont {Calvo}}, \bibinfo
  {author} {\bibfnamefont {A.}~\bibnamefont {Catalano}}, \bibinfo {author}
  {\bibfnamefont {A.}~\bibnamefont {Gomez}}, \bibinfo {author} {\bibfnamefont
  {J.}~\bibnamefont {Goupy}}, \bibinfo {author} {\bibfnamefont
  {T.}~\bibnamefont {Grenet}}, \bibinfo {author} {\bibfnamefont
  {L.}~\bibnamefont {Gr{\"u}nhaupt}}, \bibinfo {author} {\bibfnamefont
  {T.}~\bibnamefont {Klein}}, \bibinfo {author} {\bibfnamefont
  {U.}~\bibnamefont {{von L{\"u}pke}}}, \bibinfo {author} {\bibfnamefont
  {N.}~\bibnamefont {Maleeva}}, \bibinfo {author} {\bibfnamefont
  {F.}~\bibnamefont {Valenti}}, \bibinfo {author} {\bibfnamefont
  {A.}~\bibnamefont {Monfardini}}, \bibinfo {author} {\bibfnamefont {I.~M.}\
  \bibnamefont {Pop}}, \ and\ \bibinfo {author} {\bibfnamefont
  {F.}~\bibnamefont {Levy-Bertrand}},\ }\href@noop {} {\bibfield  {journal}
  {\bibinfo  {journal} {in preparation}\ ,\ } (\bibinfo {year}
  {2018})}\BibitemShut {NoStop}%
\bibitem [{\citenamefont {Deutscher}\ \emph {et~al.}(1973)\citenamefont
  {Deutscher}, \citenamefont {Fenichel}, \citenamefont {Gershenson},
  \citenamefont {Gr{\"u}nbaum},\ and\ \citenamefont
  {Ovadyahu}}]{deutscher_transition_1973}%
  \BibitemOpen
  \bibfield  {author} {\bibinfo {author} {\bibfnamefont {G.}~\bibnamefont
  {Deutscher}}, \bibinfo {author} {\bibfnamefont {H.}~\bibnamefont {Fenichel}},
  \bibinfo {author} {\bibfnamefont {M.}~\bibnamefont {Gershenson}}, \bibinfo
  {author} {\bibfnamefont {E.}~\bibnamefont {Gr{\"u}nbaum}}, \ and\ \bibinfo
  {author} {\bibfnamefont {Z.}~\bibnamefont {Ovadyahu}},\ }\href {\doibase
  10.1007/BF00655256} {\bibfield  {journal} {\bibinfo  {journal} {J. Low Temp.
  Phys.}\ }\textbf {\bibinfo {volume} {10}},\ \bibinfo {pages} {231} (\bibinfo
  {year} {1973})}\BibitemShut {NoStop}%
\bibitem [{\citenamefont {Wang}\ \emph {et~al.}(2015)\citenamefont {Wang},
  \citenamefont {Axline}, \citenamefont {Gao}, \citenamefont {Brecht},
  \citenamefont {Chu}, \citenamefont {Frunzio}, \citenamefont {Devoret},\ and\
  \citenamefont {Schoelkopf}}]{wang_surface_2015}%
  \BibitemOpen
  \bibfield  {author} {\bibinfo {author} {\bibfnamefont {C.}~\bibnamefont
  {Wang}}, \bibinfo {author} {\bibfnamefont {C.}~\bibnamefont {Axline}},
  \bibinfo {author} {\bibfnamefont {Y.~Y.}\ \bibnamefont {Gao}}, \bibinfo
  {author} {\bibfnamefont {T.}~\bibnamefont {Brecht}}, \bibinfo {author}
  {\bibfnamefont {Y.}~\bibnamefont {Chu}}, \bibinfo {author} {\bibfnamefont
  {L.}~\bibnamefont {Frunzio}}, \bibinfo {author} {\bibfnamefont {M.~H.}\
  \bibnamefont {Devoret}}, \ and\ \bibinfo {author} {\bibfnamefont {R.~J.}\
  \bibnamefont {Schoelkopf}},\ }\href {\doibase 10.1063/1.4934486} {\bibfield
  {journal} {\bibinfo  {journal} {Appl. Phys. Lett.}\ }\textbf {\bibinfo
  {volume} {107}},\ \bibinfo {pages} {162601} (\bibinfo {year}
  {2015})}\BibitemShut {NoStop}%
\bibitem [{\citenamefont {Buckel}\ and\ \citenamefont
  {Hilsch}(1954)}]{buckel_einfluss_1954}%
  \BibitemOpen
  \bibfield  {author} {\bibinfo {author} {\bibfnamefont {W.}~\bibnamefont
  {Buckel}}\ and\ \bibinfo {author} {\bibfnamefont {R.}~\bibnamefont
  {Hilsch}},\ }\href {\doibase 10.1007/BF01337903} {\bibfield  {journal}
  {\bibinfo  {journal} {Zeitschrift f{\"u}r Physik}\ }\textbf {\bibinfo
  {volume} {138}},\ \bibinfo {pages} {109} (\bibinfo {year}
  {1954})}\BibitemShut {NoStop}%
\bibitem [{\citenamefont {Cohen}\ and\ \citenamefont
  {Abeles}(1968)}]{cohen_superconductivity_1968}%
  \BibitemOpen
  \bibfield  {author} {\bibinfo {author} {\bibfnamefont {R.~W.}\ \bibnamefont
  {Cohen}}\ and\ \bibinfo {author} {\bibfnamefont {B.}~\bibnamefont {Abeles}},\
  }\href {\doibase 10.1103/PhysRev.168.444} {\bibfield  {journal} {\bibinfo
  {journal} {Phys. Rev.}\ }\textbf {\bibinfo {volume} {168}},\ \bibinfo {pages}
  {444} (\bibinfo {year} {1968})}\BibitemShut {NoStop}%
\bibitem [{\citenamefont {Rotzinger}\ \emph {et~al.}(2017)\citenamefont
  {Rotzinger}, \citenamefont {Skacel}, \citenamefont {Pfirrmann}, \citenamefont
  {Voss}, \citenamefont {M{\"u}nzberg}, \citenamefont {Probst}, \citenamefont
  {Bushev}, \citenamefont {Weides}, \citenamefont {Ustinov},\ and\
  \citenamefont {Mooij}}]{rotzinger_aluminium-oxide_2017}%
  \BibitemOpen
  \bibfield  {author} {\bibinfo {author} {\bibfnamefont {H.}~\bibnamefont
  {Rotzinger}}, \bibinfo {author} {\bibfnamefont {S.~T.}\ \bibnamefont
  {Skacel}}, \bibinfo {author} {\bibfnamefont {M.}~\bibnamefont {Pfirrmann}},
  \bibinfo {author} {\bibfnamefont {J.~N.}\ \bibnamefont {Voss}}, \bibinfo
  {author} {\bibfnamefont {J.}~\bibnamefont {M{\"u}nzberg}}, \bibinfo {author}
  {\bibfnamefont {S.}~\bibnamefont {Probst}}, \bibinfo {author} {\bibfnamefont
  {P.}~\bibnamefont {Bushev}}, \bibinfo {author} {\bibfnamefont {M.~P.}\
  \bibnamefont {Weides}}, \bibinfo {author} {\bibfnamefont {A.~V.}\
  \bibnamefont {Ustinov}}, \ and\ \bibinfo {author} {\bibfnamefont {J.~E.}\
  \bibnamefont {Mooij}},\ }\href {\doibase 10.1088/0953-2048/30/2/025002}
  {\bibfield  {journal} {\bibinfo  {journal} {Supercond. Sci. Technol.}\
  }\textbf {\bibinfo {volume} {30}},\ \bibinfo {pages} {025002} (\bibinfo
  {year} {2017})}\BibitemShut {NoStop}%
\bibitem [{\citenamefont
  {Manucharyan}(2012)}]{manucharyan_superinductance_2012}%
  \BibitemOpen
  \bibfield  {author} {\bibinfo {author} {\bibfnamefont {V.~E.}\ \bibnamefont
  {Manucharyan}},\ }\emph {\bibinfo {title} {Superinductance}},\ \href@noop {}
  {\bibinfo {type} {Dissertation}},\ \bibinfo  {school} {Yale University},
  \bibinfo {address} {New Haven, CT} (\bibinfo {year} {2012})\BibitemShut
  {NoStop}%
\bibitem [{\citenamefont {Masluk}\ \emph {et~al.}(2012)\citenamefont {Masluk},
  \citenamefont {Pop}, \citenamefont {Kamal}, \citenamefont {Minev},\ and\
  \citenamefont {Devoret}}]{masluk_microwave_2012}%
  \BibitemOpen
  \bibfield  {author} {\bibinfo {author} {\bibfnamefont {N.~A.}\ \bibnamefont
  {Masluk}}, \bibinfo {author} {\bibfnamefont {I.~M.}\ \bibnamefont {Pop}},
  \bibinfo {author} {\bibfnamefont {A.}~\bibnamefont {Kamal}}, \bibinfo
  {author} {\bibfnamefont {Z.~K.}\ \bibnamefont {Minev}}, \ and\ \bibinfo
  {author} {\bibfnamefont {M.~H.}\ \bibnamefont {Devoret}},\ }\href {\doibase
  10.1103/PhysRevLett.109.137002} {\bibfield  {journal} {\bibinfo  {journal}
  {Phys. Rev. Lett.}\ }\textbf {\bibinfo {volume} {109}},\  (\bibinfo {year}
  {2012})}\BibitemShut {NoStop}%
\bibitem [{\citenamefont {Bell}\ \emph {et~al.}(2012)\citenamefont {Bell},
  \citenamefont {Sadovskyy}, \citenamefont {Ioffe}, \citenamefont {Kitaev},\
  and\ \citenamefont {Gershenson}}]{bell_quantum_2012}%
  \BibitemOpen
  \bibfield  {author} {\bibinfo {author} {\bibfnamefont {M.~T.}\ \bibnamefont
  {Bell}}, \bibinfo {author} {\bibfnamefont {I.~A.}\ \bibnamefont {Sadovskyy}},
  \bibinfo {author} {\bibfnamefont {L.~B.}\ \bibnamefont {Ioffe}}, \bibinfo
  {author} {\bibfnamefont {A.~Y.}\ \bibnamefont {Kitaev}}, \ and\ \bibinfo
  {author} {\bibfnamefont {M.~E.}\ \bibnamefont {Gershenson}},\ }\href
  {\doibase 10.1103/PhysRevLett.109.137003} {\bibfield  {journal} {\bibinfo
  {journal} {Phys. Rev. Lett.}\ }\textbf {\bibinfo {volume} {109}},\  (\bibinfo
  {year} {2012})}\BibitemShut {NoStop}%
\bibitem [{\citenamefont {Vissers}\ \emph {et~al.}(2010)\citenamefont
  {Vissers}, \citenamefont {Gao}, \citenamefont {Wisbey}, \citenamefont {Hite},
  \citenamefont {Tsuei}, \citenamefont {Corcoles}, \citenamefont {Steffen},\
  and\ \citenamefont {Pappas}}]{vissers_low_2010}%
  \BibitemOpen
  \bibfield  {author} {\bibinfo {author} {\bibfnamefont {M.~R.}\ \bibnamefont
  {Vissers}}, \bibinfo {author} {\bibfnamefont {J.}~\bibnamefont {Gao}},
  \bibinfo {author} {\bibfnamefont {D.~S.}\ \bibnamefont {Wisbey}}, \bibinfo
  {author} {\bibfnamefont {D.~A.}\ \bibnamefont {Hite}}, \bibinfo {author}
  {\bibfnamefont {C.~C.}\ \bibnamefont {Tsuei}}, \bibinfo {author}
  {\bibfnamefont {A.~D.}\ \bibnamefont {Corcoles}}, \bibinfo {author}
  {\bibfnamefont {M.}~\bibnamefont {Steffen}}, \ and\ \bibinfo {author}
  {\bibfnamefont {D.~P.}\ \bibnamefont {Pappas}},\ }\href {\doibase
  10.1063/1.3517252} {\bibfield  {journal} {\bibinfo  {journal} {Appl. Phys.
  Lett.}\ }\textbf {\bibinfo {volume} {97}},\ \bibinfo {pages} {232509}
  (\bibinfo {year} {2010})}\BibitemShut {NoStop}%
\bibitem [{\citenamefont {Leduc}\ \emph {et~al.}(2010)\citenamefont {Leduc},
  \citenamefont {Bumble}, \citenamefont {Day}, \citenamefont {Eom},
  \citenamefont {Gao}, \citenamefont {Golwala}, \citenamefont {Mazin},
  \citenamefont {McHugh}, \citenamefont {Merrill}, \citenamefont {Moore},
  \citenamefont {Noroozian}, \citenamefont {Turner},\ and\ \citenamefont
  {Zmuidzinas}}]{leduc_titanium_2010}%
  \BibitemOpen
  \bibfield  {author} {\bibinfo {author} {\bibfnamefont {H.~G.}\ \bibnamefont
  {Leduc}}, \bibinfo {author} {\bibfnamefont {B.}~\bibnamefont {Bumble}},
  \bibinfo {author} {\bibfnamefont {P.~K.}\ \bibnamefont {Day}}, \bibinfo
  {author} {\bibfnamefont {B.~H.}\ \bibnamefont {Eom}}, \bibinfo {author}
  {\bibfnamefont {J.}~\bibnamefont {Gao}}, \bibinfo {author} {\bibfnamefont
  {S.}~\bibnamefont {Golwala}}, \bibinfo {author} {\bibfnamefont {B.~A.}\
  \bibnamefont {Mazin}}, \bibinfo {author} {\bibfnamefont {S.}~\bibnamefont
  {McHugh}}, \bibinfo {author} {\bibfnamefont {A.}~\bibnamefont {Merrill}},
  \bibinfo {author} {\bibfnamefont {D.~C.}\ \bibnamefont {Moore}}, \bibinfo
  {author} {\bibfnamefont {O.}~\bibnamefont {Noroozian}}, \bibinfo {author}
  {\bibfnamefont {A.~D.}\ \bibnamefont {Turner}}, \ and\ \bibinfo {author}
  {\bibfnamefont {J.}~\bibnamefont {Zmuidzinas}},\ }\href {\doibase
  10.1063/1.3480420} {\bibfield  {journal} {\bibinfo  {journal} {Appl. Phys.
  Lett.}\ }\textbf {\bibinfo {volume} {97}},\ \bibinfo {pages} {102509}
  (\bibinfo {year} {2010})}\BibitemShut {NoStop}%
\bibitem [{\citenamefont {Swenson}\ \emph {et~al.}(2013)\citenamefont
  {Swenson}, \citenamefont {Day}, \citenamefont {Eom}, \citenamefont {Leduc},
  \citenamefont {Llombart}, \citenamefont {McKenney}, \citenamefont
  {Noroozian},\ and\ \citenamefont {Zmuidzinas}}]{swenson_operation_2013}%
  \BibitemOpen
  \bibfield  {author} {\bibinfo {author} {\bibfnamefont {L.~J.}\ \bibnamefont
  {Swenson}}, \bibinfo {author} {\bibfnamefont {P.~K.}\ \bibnamefont {Day}},
  \bibinfo {author} {\bibfnamefont {B.~H.}\ \bibnamefont {Eom}}, \bibinfo
  {author} {\bibfnamefont {H.~G.}\ \bibnamefont {Leduc}}, \bibinfo {author}
  {\bibfnamefont {N.}~\bibnamefont {Llombart}}, \bibinfo {author}
  {\bibfnamefont {C.~M.}\ \bibnamefont {McKenney}}, \bibinfo {author}
  {\bibfnamefont {O.}~\bibnamefont {Noroozian}}, \ and\ \bibinfo {author}
  {\bibfnamefont {J.}~\bibnamefont {Zmuidzinas}},\ }\href {\doibase
  10.1063/1.4794808} {\bibfield  {journal} {\bibinfo  {journal} {J. Appl.
  Phys.}\ }\textbf {\bibinfo {volume} {113}},\ \bibinfo {pages} {104501}
  (\bibinfo {year} {2013})}\BibitemShut {NoStop}%
\bibitem [{\citenamefont {Barends}\ \emph {et~al.}(2010)\citenamefont
  {Barends}, \citenamefont {Vercruyssen}, \citenamefont {Endo}, \citenamefont
  {{de Visser}}, \citenamefont {Zijlstra}, \citenamefont {Klapwijk},\ and\
  \citenamefont {Baselmans}}]{barends_reduced_2010}%
  \BibitemOpen
  \bibfield  {author} {\bibinfo {author} {\bibfnamefont {R.}~\bibnamefont
  {Barends}}, \bibinfo {author} {\bibfnamefont {N.}~\bibnamefont
  {Vercruyssen}}, \bibinfo {author} {\bibfnamefont {A.}~\bibnamefont {Endo}},
  \bibinfo {author} {\bibfnamefont {P.~J.}\ \bibnamefont {{de Visser}}},
  \bibinfo {author} {\bibfnamefont {T.}~\bibnamefont {Zijlstra}}, \bibinfo
  {author} {\bibfnamefont {T.~M.}\ \bibnamefont {Klapwijk}}, \ and\ \bibinfo
  {author} {\bibfnamefont {J.~J.~A.}\ \bibnamefont {Baselmans}},\ }\href
  {\doibase 10.1063/1.3467052} {\bibfield  {journal} {\bibinfo  {journal}
  {Appl. Phys. Lett.}\ }\textbf {\bibinfo {volume} {97}},\ \bibinfo {pages}
  {033507} (\bibinfo {year} {2010})}\BibitemShut {NoStop}%
\bibitem [{\citenamefont {Samkharadze}\ \emph {et~al.}(2016)\citenamefont
  {Samkharadze}, \citenamefont {Bruno}, \citenamefont {Scarlino}, \citenamefont
  {Zheng}, \citenamefont {DiVincenzo}, \citenamefont {DiCarlo},\ and\
  \citenamefont {Vandersypen}}]{samkharadze_high_kinetic_inductance_2016}%
  \BibitemOpen
  \bibfield  {author} {\bibinfo {author} {\bibfnamefont {N.}~\bibnamefont
  {Samkharadze}}, \bibinfo {author} {\bibfnamefont {A.}~\bibnamefont {Bruno}},
  \bibinfo {author} {\bibfnamefont {P.}~\bibnamefont {Scarlino}}, \bibinfo
  {author} {\bibfnamefont {G.}~\bibnamefont {Zheng}}, \bibinfo {author}
  {\bibfnamefont {D.~P.}\ \bibnamefont {DiVincenzo}}, \bibinfo {author}
  {\bibfnamefont {L.}~\bibnamefont {DiCarlo}}, \ and\ \bibinfo {author}
  {\bibfnamefont {L.~M.~K.}\ \bibnamefont {Vandersypen}},\ }\href {\doibase
  10.1103/PhysRevApplied.5.044004} {\bibfield  {journal} {\bibinfo  {journal}
  {Phys. Rev. Appl}\ }\textbf {\bibinfo {volume} {5}},\  (\bibinfo {year}
  {2016})}\BibitemShut {NoStop}%
\bibitem [{\citenamefont {Grabovskij}\ \emph {et~al.}(2008)\citenamefont
  {Grabovskij}, \citenamefont {Swenson}, \citenamefont {Buisson}, \citenamefont
  {Hoffmann}, \citenamefont {Monfardini},\ and\ \citenamefont
  {Vill{\'e}gier}}]{grabovskij_textitsitu_2008}%
  \BibitemOpen
  \bibfield  {author} {\bibinfo {author} {\bibfnamefont {G.~J.}\ \bibnamefont
  {Grabovskij}}, \bibinfo {author} {\bibfnamefont {L.~J.}\ \bibnamefont
  {Swenson}}, \bibinfo {author} {\bibfnamefont {O.}~\bibnamefont {Buisson}},
  \bibinfo {author} {\bibfnamefont {C.}~\bibnamefont {Hoffmann}}, \bibinfo
  {author} {\bibfnamefont {A.}~\bibnamefont {Monfardini}}, \ and\ \bibinfo
  {author} {\bibfnamefont {J.-C.}\ \bibnamefont {Vill{\'e}gier}},\ }\href
  {\doibase 10.1063/1.2996263} {\bibfield  {journal} {\bibinfo  {journal}
  {Appl. Phys. Lett.}\ }\textbf {\bibinfo {volume} {93}},\ \bibinfo {pages}
  {134102} (\bibinfo {year} {2008})}\BibitemShut {NoStop}%
\bibitem [{\citenamefont {Luomahaara}\ \emph {et~al.}(2014)\citenamefont
  {Luomahaara}, \citenamefont {Vesterinen}, \citenamefont {Gr{\"o}nberg},\ and\
  \citenamefont {Hassel}}]{luomahaara_kinetic_2014}%
  \BibitemOpen
  \bibfield  {author} {\bibinfo {author} {\bibfnamefont {J.}~\bibnamefont
  {Luomahaara}}, \bibinfo {author} {\bibfnamefont {V.}~\bibnamefont
  {Vesterinen}}, \bibinfo {author} {\bibfnamefont {L.}~\bibnamefont
  {Gr{\"o}nberg}}, \ and\ \bibinfo {author} {\bibfnamefont {J.}~\bibnamefont
  {Hassel}},\ }\href {\doibase 10.1038/ncomms5872} {\bibfield  {journal}
  {\bibinfo  {journal} {Nat. Commun.}\ }\textbf {\bibinfo {volume} {5}},\
  \bibinfo {pages} {4872} (\bibinfo {year} {2014})}\BibitemShut {NoStop}%
\bibitem [{\citenamefont {Feigel'man}\ and\ \citenamefont
  {Ioffe}(2018)}]{feigelman_microwave_2018}%
  \BibitemOpen
  \bibfield  {author} {\bibinfo {author} {\bibfnamefont {M.~V.}\ \bibnamefont
  {Feigel'man}}\ and\ \bibinfo {author} {\bibfnamefont {L.~B.}\ \bibnamefont
  {Ioffe}},\ }\href {\doibase 10.1103/PhysRevLett.120.037004} {\bibfield
  {journal} {\bibinfo  {journal} {Phys. Rev. Lett.}\ }\textbf {\bibinfo
  {volume} {120}},\  (\bibinfo {year} {2018})}\BibitemShut {NoStop}%
\bibitem [{\citenamefont {Gr{\"u}nhaupt}\ \emph {et~al.}(2017)\citenamefont
  {Gr{\"u}nhaupt}, \citenamefont {{von L{\"u}pke}}, \citenamefont {Gusenkova},
  \citenamefont {Skacel}, \citenamefont {Maleeva}, \citenamefont {Schl{\"o}r},
  \citenamefont {Bilmes}, \citenamefont {Rotzinger}, \citenamefont {Ustinov},
  \citenamefont {Weides},\ and\ \citenamefont {Pop}}]{grunhaupt_argon_2017}%
  \BibitemOpen
  \bibfield  {author} {\bibinfo {author} {\bibfnamefont {L.}~\bibnamefont
  {Gr{\"u}nhaupt}}, \bibinfo {author} {\bibfnamefont {U.}~\bibnamefont {{von
  L{\"u}pke}}}, \bibinfo {author} {\bibfnamefont {D.}~\bibnamefont
  {Gusenkova}}, \bibinfo {author} {\bibfnamefont {S.~T.}\ \bibnamefont
  {Skacel}}, \bibinfo {author} {\bibfnamefont {N.}~\bibnamefont {Maleeva}},
  \bibinfo {author} {\bibfnamefont {S.}~\bibnamefont {Schl{\"o}r}}, \bibinfo
  {author} {\bibfnamefont {A.}~\bibnamefont {Bilmes}}, \bibinfo {author}
  {\bibfnamefont {H.}~\bibnamefont {Rotzinger}}, \bibinfo {author}
  {\bibfnamefont {A.~V.}\ \bibnamefont {Ustinov}}, \bibinfo {author}
  {\bibfnamefont {M.}~\bibnamefont {Weides}}, \ and\ \bibinfo {author}
  {\bibfnamefont {I.~M.}\ \bibnamefont {Pop}},\ }\href {\doibase
  10.1063/1.4990491} {\bibfield  {journal} {\bibinfo  {journal} {Appl. Phys.
  Lett.}\ }\textbf {\bibinfo {volume} {111}},\ \bibinfo {pages} {072601}
  (\bibinfo {year} {2017})}\BibitemShut {NoStop}%
\bibitem [{\citenamefont {Wei{\ss}l}\ \emph {et~al.}(2015)\citenamefont
  {Wei{\ss}l}, \citenamefont {K{\"u}ng}, \citenamefont {Dumur}, \citenamefont
  {Feofanov}, \citenamefont {Matei}, \citenamefont {Naud}, \citenamefont
  {Buisson}, \citenamefont {Hekking},\ and\ \citenamefont
  {Guichard}}]{weisl_kerr_2015}%
  \BibitemOpen
  \bibfield  {author} {\bibinfo {author} {\bibfnamefont {T.}~\bibnamefont
  {Wei{\ss}l}}, \bibinfo {author} {\bibfnamefont {B.}~\bibnamefont {K{\"u}ng}},
  \bibinfo {author} {\bibfnamefont {E.}~\bibnamefont {Dumur}}, \bibinfo
  {author} {\bibfnamefont {A.~K.}\ \bibnamefont {Feofanov}}, \bibinfo {author}
  {\bibfnamefont {I.}~\bibnamefont {Matei}}, \bibinfo {author} {\bibfnamefont
  {C.}~\bibnamefont {Naud}}, \bibinfo {author} {\bibfnamefont {O.}~\bibnamefont
  {Buisson}}, \bibinfo {author} {\bibfnamefont {F.~W.~J.}\ \bibnamefont
  {Hekking}}, \ and\ \bibinfo {author} {\bibfnamefont {W.}~\bibnamefont
  {Guichard}},\ }\href {\doibase 10.1103/PhysRevB.92.104508} {\bibfield
  {journal} {\bibinfo  {journal} {Phys. Rev. B}\ }\textbf {\bibinfo {volume}
  {92}},\  (\bibinfo {year} {2015})}\BibitemShut {NoStop}%
\bibitem [{\citenamefont {Maleeva}\ \emph {et~al.}(2018)\citenamefont
  {Maleeva}, \citenamefont {Gr{\"u}nhaupt}, \citenamefont {Klein},
  \citenamefont {Levy-Bertrand}, \citenamefont {Dupr{\'e}}, \citenamefont
  {Calvo}, \citenamefont {Valenti}, \citenamefont {Winkel}, \citenamefont
  {Friedrich}, \citenamefont {Wernsdorfer}, \citenamefont {Ustinov},
  \citenamefont {Rotzinger}, \citenamefont {Monfardini}, \citenamefont
  {Fistul},\ and\ \citenamefont {Pop}}]{maleeva_circuit_2018}%
  \BibitemOpen
  \bibfield  {author} {\bibinfo {author} {\bibfnamefont {N.}~\bibnamefont
  {Maleeva}}, \bibinfo {author} {\bibfnamefont {L.}~\bibnamefont
  {Gr{\"u}nhaupt}}, \bibinfo {author} {\bibfnamefont {T.}~\bibnamefont
  {Klein}}, \bibinfo {author} {\bibfnamefont {F.}~\bibnamefont
  {Levy-Bertrand}}, \bibinfo {author} {\bibfnamefont {O.}~\bibnamefont
  {Dupr{\'e}}}, \bibinfo {author} {\bibfnamefont {M.}~\bibnamefont {Calvo}},
  \bibinfo {author} {\bibfnamefont {F.}~\bibnamefont {Valenti}}, \bibinfo
  {author} {\bibfnamefont {P.}~\bibnamefont {Winkel}}, \bibinfo {author}
  {\bibfnamefont {F.}~\bibnamefont {Friedrich}}, \bibinfo {author}
  {\bibfnamefont {W.}~\bibnamefont {Wernsdorfer}}, \bibinfo {author}
  {\bibfnamefont {A.~V.}\ \bibnamefont {Ustinov}}, \bibinfo {author}
  {\bibfnamefont {H.}~\bibnamefont {Rotzinger}}, \bibinfo {author}
  {\bibfnamefont {A.}~\bibnamefont {Monfardini}}, \bibinfo {author}
  {\bibfnamefont {M.~V.}\ \bibnamefont {Fistul}}, \ and\ \bibinfo {author}
  {\bibfnamefont {I.~M.}\ \bibnamefont {Pop}},\ }\href
  {https://arxiv.org/abs/1802.01859} {\bibfield  {journal} {\bibinfo  {journal}
  {arXiv:1802.01859 [cond-mat.supr-con]}\ ,\ } (\bibinfo {year}
  {2018})}\BibitemShut {NoStop}%
\bibitem [{\citenamefont {Hunklinger}\ \emph {et~al.}(1972)\citenamefont
  {Hunklinger}, \citenamefont {Arnold}, \citenamefont {Stein}, \citenamefont
  {Nava},\ and\ \citenamefont {Dransfeld}}]{hunklinger_saturation_1972}%
  \BibitemOpen
  \bibfield  {author} {\bibinfo {author} {\bibfnamefont {S.}~\bibnamefont
  {Hunklinger}}, \bibinfo {author} {\bibfnamefont {W.}~\bibnamefont {Arnold}},
  \bibinfo {author} {\bibfnamefont {S.}~\bibnamefont {Stein}}, \bibinfo
  {author} {\bibfnamefont {R.}~\bibnamefont {Nava}}, \ and\ \bibinfo {author}
  {\bibfnamefont {K.}~\bibnamefont {Dransfeld}},\ }\href {\doibase
  10.1016/0375-9601(72)90884-5} {\bibfield  {journal} {\bibinfo  {journal}
  {Phys. Lett. A}\ }\textbf {\bibinfo {volume} {42}},\ \bibinfo {pages} {253}
  (\bibinfo {year} {1972})}\BibitemShut {NoStop}%
\bibitem [{\citenamefont {Golding}\ \emph {et~al.}(1973)\citenamefont
  {Golding}, \citenamefont {Graebner}, \citenamefont {Halperin},\ and\
  \citenamefont {Schutz}}]{golding_nonlinear_1973}%
  \BibitemOpen
  \bibfield  {author} {\bibinfo {author} {\bibfnamefont {B.}~\bibnamefont
  {Golding}}, \bibinfo {author} {\bibfnamefont {J.~E.}\ \bibnamefont
  {Graebner}}, \bibinfo {author} {\bibfnamefont {B.~I.}\ \bibnamefont
  {Halperin}}, \ and\ \bibinfo {author} {\bibfnamefont {R.~J.}\ \bibnamefont
  {Schutz}},\ }\href {\doibase 10.1103/PhysRevLett.30.223} {\bibfield
  {journal} {\bibinfo  {journal} {Phys. Rev. Lett.}\ }\textbf {\bibinfo
  {volume} {30}},\ \bibinfo {pages} {223} (\bibinfo {year} {1973})}\BibitemShut
  {NoStop}%
\bibitem [{\citenamefont {Wenner}\ \emph {et~al.}(2011)\citenamefont {Wenner},
  \citenamefont {Barends}, \citenamefont {Bialczak}, \citenamefont {Chen},
  \citenamefont {Kelly}, \citenamefont {Lucero}, \citenamefont {Mariantoni},
  \citenamefont {Megrant}, \citenamefont {O'Malley}, \citenamefont {Sank},
  \citenamefont {Vainsencher}, \citenamefont {Wang}, \citenamefont {White},
  \citenamefont {Yin}, \citenamefont {Zhao}, \citenamefont {Cleland},\ and\
  \citenamefont {Martinis}}]{wenner_surface_2011}%
  \BibitemOpen
  \bibfield  {author} {\bibinfo {author} {\bibfnamefont {J.}~\bibnamefont
  {Wenner}}, \bibinfo {author} {\bibfnamefont {R.}~\bibnamefont {Barends}},
  \bibinfo {author} {\bibfnamefont {R.~C.}\ \bibnamefont {Bialczak}}, \bibinfo
  {author} {\bibfnamefont {Y.}~\bibnamefont {Chen}}, \bibinfo {author}
  {\bibfnamefont {J.}~\bibnamefont {Kelly}}, \bibinfo {author} {\bibfnamefont
  {E.}~\bibnamefont {Lucero}}, \bibinfo {author} {\bibfnamefont
  {M.}~\bibnamefont {Mariantoni}}, \bibinfo {author} {\bibfnamefont
  {A.}~\bibnamefont {Megrant}}, \bibinfo {author} {\bibfnamefont {P.~J.~J.}\
  \bibnamefont {O'Malley}}, \bibinfo {author} {\bibfnamefont {D.}~\bibnamefont
  {Sank}}, \bibinfo {author} {\bibfnamefont {A.}~\bibnamefont {Vainsencher}},
  \bibinfo {author} {\bibfnamefont {H.}~\bibnamefont {Wang}}, \bibinfo {author}
  {\bibfnamefont {T.~C.}\ \bibnamefont {White}}, \bibinfo {author}
  {\bibfnamefont {Y.}~\bibnamefont {Yin}}, \bibinfo {author} {\bibfnamefont
  {J.}~\bibnamefont {Zhao}}, \bibinfo {author} {\bibfnamefont {A.~N.}\
  \bibnamefont {Cleland}}, \ and\ \bibinfo {author} {\bibfnamefont {J.~M.}\
  \bibnamefont {Martinis}},\ }\href {\doibase 10.1063/1.3637047} {\bibfield
  {journal} {\bibinfo  {journal} {Appl. Phys. Lett.}\ }\textbf {\bibinfo
  {volume} {99}},\ \bibinfo {pages} {113513} (\bibinfo {year}
  {2011})}\BibitemShut {NoStop}%
\bibitem [{\citenamefont {D'Addabbo}(2014)}]{daddabbo_applications_2014}%
  \BibitemOpen
  \bibfield  {author} {\bibinfo {author} {\bibfnamefont {A.}~\bibnamefont
  {D'Addabbo}},\ }\emph {\bibinfo {title} {Applications of {{Kinetic Inductance
  Detectors}} to {{Astronomy}} and {{Particle Physics}}}},\ \href
  {https://tel.archives-ouvertes.fr/tel-01368000} {\bibinfo {type} {Phd
  dissertation}},\ \bibinfo  {school} {Universit{\'e} de Grenoble} (\bibinfo
  {year} {2014})\BibitemShut {NoStop}%
\bibitem [{\citenamefont {Swenson}\ \emph {et~al.}(2010)\citenamefont
  {Swenson}, \citenamefont {Cruciani}, \citenamefont {Benoit}, \citenamefont
  {Roesch}, \citenamefont {Yung}, \citenamefont {Bideaud},\ and\ \citenamefont
  {Monfardini}}]{swenson_high-speed_2010}%
  \BibitemOpen
  \bibfield  {author} {\bibinfo {author} {\bibfnamefont {L.~J.}\ \bibnamefont
  {Swenson}}, \bibinfo {author} {\bibfnamefont {A.}~\bibnamefont {Cruciani}},
  \bibinfo {author} {\bibfnamefont {A.}~\bibnamefont {Benoit}}, \bibinfo
  {author} {\bibfnamefont {M.}~\bibnamefont {Roesch}}, \bibinfo {author}
  {\bibfnamefont {C.~S.}\ \bibnamefont {Yung}}, \bibinfo {author}
  {\bibfnamefont {A.}~\bibnamefont {Bideaud}}, \ and\ \bibinfo {author}
  {\bibfnamefont {A.}~\bibnamefont {Monfardini}},\ }\href {\doibase
  10.1063/1.3459142} {\bibfield  {journal} {\bibinfo  {journal} {Appl. Phys.
  Lett.}\ }\textbf {\bibinfo {volume} {96}},\ \bibinfo {pages} {263511}
  (\bibinfo {year} {2010})}\BibitemShut {NoStop}%
\bibitem [{\citenamefont {Moore}\ \emph {et~al.}(2012)\citenamefont {Moore},
  \citenamefont {Golwala}, \citenamefont {Bumble}, \citenamefont {Cornell},
  \citenamefont {Day}, \citenamefont {LeDuc},\ and\ \citenamefont
  {Zmuidzinas}}]{moore_position_2012}%
  \BibitemOpen
  \bibfield  {author} {\bibinfo {author} {\bibfnamefont {D.~C.}\ \bibnamefont
  {Moore}}, \bibinfo {author} {\bibfnamefont {S.~R.}\ \bibnamefont {Golwala}},
  \bibinfo {author} {\bibfnamefont {B.}~\bibnamefont {Bumble}}, \bibinfo
  {author} {\bibfnamefont {B.}~\bibnamefont {Cornell}}, \bibinfo {author}
  {\bibfnamefont {P.~K.}\ \bibnamefont {Day}}, \bibinfo {author} {\bibfnamefont
  {H.~G.}\ \bibnamefont {LeDuc}}, \ and\ \bibinfo {author} {\bibfnamefont
  {J.}~\bibnamefont {Zmuidzinas}},\ }\href {\doibase 10.1063/1.4726279}
  {\bibfield  {journal} {\bibinfo  {journal} {Appl. Phys. Lett.}\ }\textbf
  {\bibinfo {volume} {100}},\ \bibinfo {pages} {232601} (\bibinfo {year}
  {2012})}\BibitemShut {NoStop}%
\bibitem [{\citenamefont {Cardani}\ \emph {et~al.}(2015)\citenamefont
  {Cardani}, \citenamefont {Colantoni}, \citenamefont {Cruciani}, \citenamefont
  {Di~Domizio}, \citenamefont {Vignati}, \citenamefont {Bellini}, \citenamefont
  {Casali}, \citenamefont {Castellano}, \citenamefont {Coppolecchia},
  \citenamefont {Cosmelli},\ and\ \citenamefont {Tomei}}]{cardani_energy_2015}%
  \BibitemOpen
  \bibfield  {author} {\bibinfo {author} {\bibfnamefont {L.}~\bibnamefont
  {Cardani}}, \bibinfo {author} {\bibfnamefont {I.}~\bibnamefont {Colantoni}},
  \bibinfo {author} {\bibfnamefont {A.}~\bibnamefont {Cruciani}}, \bibinfo
  {author} {\bibfnamefont {S.}~\bibnamefont {Di~Domizio}}, \bibinfo {author}
  {\bibfnamefont {M.}~\bibnamefont {Vignati}}, \bibinfo {author} {\bibfnamefont
  {F.}~\bibnamefont {Bellini}}, \bibinfo {author} {\bibfnamefont
  {N.}~\bibnamefont {Casali}}, \bibinfo {author} {\bibfnamefont {M.~G.}\
  \bibnamefont {Castellano}}, \bibinfo {author} {\bibfnamefont
  {A.}~\bibnamefont {Coppolecchia}}, \bibinfo {author} {\bibfnamefont
  {C.}~\bibnamefont {Cosmelli}}, \ and\ \bibinfo {author} {\bibfnamefont
  {C.}~\bibnamefont {Tomei}},\ }\href {\doibase 10.1063/1.4929977} {\bibfield
  {journal} {\bibinfo  {journal} {Appl. Phys. Lett.}\ }\textbf {\bibinfo
  {volume} {107}},\ \bibinfo {pages} {093508} (\bibinfo {year}
  {2015})}\BibitemShut {NoStop}%
\bibitem [{\citenamefont {{de Visser}}\ \emph {et~al.}(2014)\citenamefont {{de
  Visser}}, \citenamefont {Goldie}, \citenamefont {Diener}, \citenamefont
  {Withington}, \citenamefont {Baselmans},\ and\ \citenamefont
  {Klapwijk}}]{de_visser_evidence_2014}%
  \BibitemOpen
  \bibfield  {author} {\bibinfo {author} {\bibfnamefont {P.~J.}\ \bibnamefont
  {{de Visser}}}, \bibinfo {author} {\bibfnamefont {D.~J.}\ \bibnamefont
  {Goldie}}, \bibinfo {author} {\bibfnamefont {P.}~\bibnamefont {Diener}},
  \bibinfo {author} {\bibfnamefont {S.}~\bibnamefont {Withington}}, \bibinfo
  {author} {\bibfnamefont {J.~J.~A.}\ \bibnamefont {Baselmans}}, \ and\
  \bibinfo {author} {\bibfnamefont {T.~M.}\ \bibnamefont {Klapwijk}},\ }\href
  {\doibase 10.1103/PhysRevLett.112.047004} {\bibfield  {journal} {\bibinfo
  {journal} {Phys. Rev. Lett.}\ }\textbf {\bibinfo {volume} {112}},\  (\bibinfo
  {year} {2014})}\BibitemShut {NoStop}%
\bibitem [{\citenamefont {Bespalov}\ \emph {et~al.}(2016)\citenamefont
  {Bespalov}, \citenamefont {Houzet}, \citenamefont {Meyer},\ and\
  \citenamefont {Nazarov}}]{bespalov_theoretical_2016}%
  \BibitemOpen
  \bibfield  {author} {\bibinfo {author} {\bibfnamefont {A.}~\bibnamefont
  {Bespalov}}, \bibinfo {author} {\bibfnamefont {M.}~\bibnamefont {Houzet}},
  \bibinfo {author} {\bibfnamefont {J.~S.}\ \bibnamefont {Meyer}}, \ and\
  \bibinfo {author} {\bibfnamefont {Y.~V.}\ \bibnamefont {Nazarov}},\ }\href
  {\doibase 10.1103/PhysRevLett.117.117002} {\bibfield  {journal} {\bibinfo
  {journal} {Phys. Rev. Lett.}\ }\textbf {\bibinfo {volume} {117}},\  (\bibinfo
  {year} {2016})}\BibitemShut {NoStop}%
\bibitem [{\citenamefont {Rothwarf}\ and\ \citenamefont
  {Taylor}(1967)}]{rothwarf_measurement_1967}%
  \BibitemOpen
  \bibfield  {author} {\bibinfo {author} {\bibfnamefont {A.}~\bibnamefont
  {Rothwarf}}\ and\ \bibinfo {author} {\bibfnamefont {B.~N.}\ \bibnamefont
  {Taylor}},\ }\href {\doibase 10.1103/PhysRevLett.19.27} {\bibfield  {journal}
  {\bibinfo  {journal} {Phys. Rev. Lett.}\ }\textbf {\bibinfo {volume} {19}},\
  \bibinfo {pages} {27} (\bibinfo {year} {1967})}\BibitemShut {NoStop}%
\bibitem [{\citenamefont {Vool}\ \emph {et~al.}(2014)\citenamefont {Vool},
  \citenamefont {Pop}, \citenamefont {Sliwa}, \citenamefont {Abdo},
  \citenamefont {Wang}, \citenamefont {Brecht}, \citenamefont {Gao},
  \citenamefont {Shankar}, \citenamefont {Hatridge}, \citenamefont {Catelani},
  \citenamefont {Mirrahimi}, \citenamefont {Frunzio}, \citenamefont
  {Schoelkopf}, \citenamefont {Glazman},\ and\ \citenamefont
  {Devoret}}]{vool_non-poissonian_2014}%
  \BibitemOpen
  \bibfield  {author} {\bibinfo {author} {\bibfnamefont {U.}~\bibnamefont
  {Vool}}, \bibinfo {author} {\bibfnamefont {I.~M.}\ \bibnamefont {Pop}},
  \bibinfo {author} {\bibfnamefont {K.}~\bibnamefont {Sliwa}}, \bibinfo
  {author} {\bibfnamefont {B.}~\bibnamefont {Abdo}}, \bibinfo {author}
  {\bibfnamefont {C.}~\bibnamefont {Wang}}, \bibinfo {author} {\bibfnamefont
  {T.}~\bibnamefont {Brecht}}, \bibinfo {author} {\bibfnamefont {Y.~Y.}\
  \bibnamefont {Gao}}, \bibinfo {author} {\bibfnamefont {S.}~\bibnamefont
  {Shankar}}, \bibinfo {author} {\bibfnamefont {M.}~\bibnamefont {Hatridge}},
  \bibinfo {author} {\bibfnamefont {G.}~\bibnamefont {Catelani}}, \bibinfo
  {author} {\bibfnamefont {M.}~\bibnamefont {Mirrahimi}}, \bibinfo {author}
  {\bibfnamefont {L.}~\bibnamefont {Frunzio}}, \bibinfo {author} {\bibfnamefont
  {R.~J.}\ \bibnamefont {Schoelkopf}}, \bibinfo {author} {\bibfnamefont
  {L.~I.}\ \bibnamefont {Glazman}}, \ and\ \bibinfo {author} {\bibfnamefont
  {M.~H.}\ \bibnamefont {Devoret}},\ }\href {\doibase
  10.1103/PhysRevLett.113.247001} {\bibfield  {journal} {\bibinfo  {journal}
  {Phys. Rev. Lett.}\ }\textbf {\bibinfo {volume} {113}},\  (\bibinfo {year}
  {2014})}\BibitemShut {NoStop}%
\bibitem [{\citenamefont {Riwar}\ \emph {et~al.}(2016)\citenamefont {Riwar},
  \citenamefont {Hosseinkhani}, \citenamefont {Burkhart}, \citenamefont {Gao},
  \citenamefont {Schoelkopf}, \citenamefont {Glazman},\ and\ \citenamefont
  {Catelani}}]{riwar_normal-metal_2016}%
  \BibitemOpen
  \bibfield  {author} {\bibinfo {author} {\bibfnamefont {R.-P.}\ \bibnamefont
  {Riwar}}, \bibinfo {author} {\bibfnamefont {A.}~\bibnamefont {Hosseinkhani}},
  \bibinfo {author} {\bibfnamefont {L.~D.}\ \bibnamefont {Burkhart}}, \bibinfo
  {author} {\bibfnamefont {Y.~Y.}\ \bibnamefont {Gao}}, \bibinfo {author}
  {\bibfnamefont {R.~J.}\ \bibnamefont {Schoelkopf}}, \bibinfo {author}
  {\bibfnamefont {L.~I.}\ \bibnamefont {Glazman}}, \ and\ \bibinfo {author}
  {\bibfnamefont {G.}~\bibnamefont {Catelani}},\ }\href {\doibase
  10.1103/PhysRevB.94.104516} {\bibfield  {journal} {\bibinfo  {journal} {Phys.
  Rev. B}\ }\textbf {\bibinfo {volume} {94}},\  (\bibinfo {year}
  {2016})}\BibitemShut {NoStop}%
\bibitem [{\citenamefont {Kaplan}\ \emph {et~al.}(1976)\citenamefont {Kaplan},
  \citenamefont {Chi}, \citenamefont {Langenberg}, \citenamefont {Chang},
  \citenamefont {Jafarey},\ and\ \citenamefont
  {Scalapino}}]{kaplan_quasiparticle_1976}%
  \BibitemOpen
  \bibfield  {author} {\bibinfo {author} {\bibfnamefont {S.~B.}\ \bibnamefont
  {Kaplan}}, \bibinfo {author} {\bibfnamefont {C.~C.}\ \bibnamefont {Chi}},
  \bibinfo {author} {\bibfnamefont {D.~N.}\ \bibnamefont {Langenberg}},
  \bibinfo {author} {\bibfnamefont {J.~J.}\ \bibnamefont {Chang}}, \bibinfo
  {author} {\bibfnamefont {S.}~\bibnamefont {Jafarey}}, \ and\ \bibinfo
  {author} {\bibfnamefont {D.~J.}\ \bibnamefont {Scalapino}},\ }\href {\doibase
  10.1103/PhysRevB.14.4854} {\bibfield  {journal} {\bibinfo  {journal} {Phys.
  Rev. B}\ }\textbf {\bibinfo {volume} {14}},\ \bibinfo {pages} {4854}
  (\bibinfo {year} {1976})}\BibitemShut {NoStop}%
\end{thebibliography}%

\clearpage 

\onecolumngrid
\renewcommand{\thefigure}{S\arabic{figure}}
\renewcommand{\thetable}{S\arabic{table}}
\renewcommand{\theequation}{S\arabic{equation}}
\setcounter{figure}{0}
\setcounter{table}{0}

\section*{Supplementary material}

\subsection*{Resonator geometries used to extract the dielectric loss tangent}

In Table \ref{suptab:1} we provide information on the metal-substrate participation ratio $p_{\mathrm{MS}}$, geometry, resonator and ground plane material of the different measured coplanar samples for the extraction of the dielectric loss tangent in Fig.~2c in the main text. $p_{\mathrm{MS}}$ is extracted from a FEM eigenmode simulation by integration of the electric field energy in a $\unit{3}{\nano\meter}$ thick volume beneath the resonator geometry following the methodology of Ref. \cite{wenner_surface_2011, wang_surface_2015}.

\begin{table}[htb]
\begin{center}
\caption{\label{suptab:1} A list of metal-substrate participation ratio $p_{\mathrm{MS}}$, resonator material, ground plane material, geometric parameters, number of measured resonators and resonant frequency.}
\begin{tabular}{cccccc}
\hline \hline 
$p_{\mathrm{MS}}$ & resonator & ground plane & geometry & no. of resonators & resonant frequency (GHz) \\ \hline %& $Q_{\mathrm{c}}$ 

 & GrAl  & GrAl& CPW, $\lambda / 4$ &  &  \\ 
$16.6 \times 10^{-3}$  & $\unit{2 \times 10^{3}}{\micro\ohm\cdot\centi\meter}$ & $\unit{2 \times 10^{3}}{\micro\ohm\cdot\centi\meter}$ & gap = $\unit{1.2}{\micro\meter}$ & 1 & 5.341 \\
 & \unit{20}{\nano\meter} & \unit{20}{\nano\meter} & strip = $\unit{2.0}{\micro\meter}$ & & \\ \hline
 
  & GrAl  & GrAl& CPW, $\lambda / 4$ &  &  \\ 
$2.9 \times 10^{-3}$  & $\unit{2 \times 10^{3}}{\micro\ohm\cdot\centi\meter}$ & $\unit{2 \times 10^{3}}{\micro\ohm\cdot\centi\meter}$ & gap = $\unit{4.8}{\micro\meter}$ & 1 & 7.447 \\
 & \unit{20}{\nano\meter} & \unit{20}{\nano\meter} & strip = $\unit{8.0}{\micro\meter}$ & & \\ \hline
 
   & GrAl  & GrAl& CPW, $\lambda / 4$ &  &  \\ 
$1.3\times 10^{-3}$  & $\unit{2 \times 10^{3}}{\micro\ohm\cdot\centi\meter}$ & $\unit{2 \times 10^{3}}{\micro\ohm\cdot\centi\meter}$ & gap = $\unit{9.6}{\micro\meter}$ & 1 & 8.186 \\
 & \unit{20}{\nano\meter} & \unit{20}{\nano\meter} & strip = $\unit{16.0}{\micro\meter}$ & & \\ \hline
 
    & GrAl  & Al & lumped element, interdigitated capacitor &  &  \\ 
$1.0\times 10^{-3}$  & $\unit{3 \times 10^{3}}{\micro\ohm\cdot\centi\meter}$ & - & capacitor gap = $\unit{12}{\micro\meter}$ & 21 & 2.167 - 2.983  \\
 & \unit{20}{\nano\meter} & \unit{20}{\nano\meter} & strip = $\unit{12}{\micro\meter}$ & & \\ \hline
 
     & GrAl  & Al & lumped element, interdigitated capacitor &  &  \\ 
$1.0\times 10^{-3}$  & $\unit{0.9 \times 10^{3}}{\micro\ohm\cdot\centi\meter}$ & - & capacitor gap = $\unit{12}{\micro\meter}$ & 21 & 2.683 - 3.767  \\
 & \unit{20}{\nano\meter} & \unit{20}{\nano\meter} & strip = $\unit{12}{\micro\meter}$ & & \\ \hline
 
\end{tabular}
\end{center}
\end{table}

\subsection*{Measurement setup and sample shielding}

\begin{figure*}[htb]
\centering
\includegraphics[scale=1]{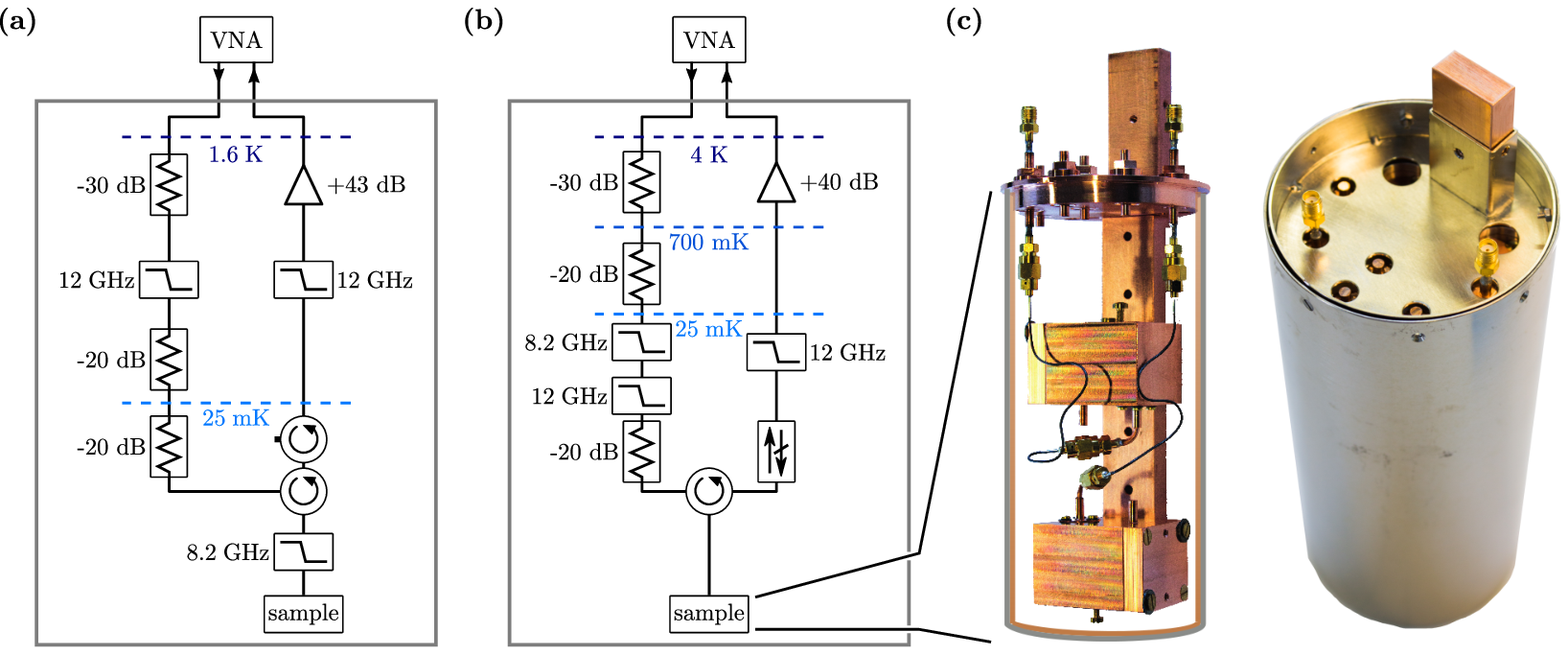}
\caption{\label{supfig:1} Schematic of the two measurement setups and illustration of the sample shielding. \textbf{(a)} Cryogenic measurement setup in a commercial dilution refrigerator with a liquid helium bath used in run \#1. \textbf{(b)} Cryogenic measurement setup in a commercial dilution refrigerator without a liquid helium bath used in runs \#2 and \#3. \textbf{(c)} Illustration of the shielding assembly. Two waveguide sample holders are mounted inside a copper/aluminum barrel, which is tightly closed. For additional magnetic shielding we enclose the barrel with a $\mu$-metal shield.}
\end{figure*}

We perform standard microwave spectroscopy measurements with a commercial vector network analyzer (VNA) on our samples. Run \#1 was measured in a commercial dilution refrigerator with a liquid helium bath. Panels (a) (run \#1) and (b) (runs \#2 and \#3) of Fig.~\ref{supfig:1} schematically show the two measurement setups. Commercial attenuators are distributed over multiple temperature stages of the cryostats to attenuate room temperature noise and successively thermalize the input lines. In total, the input signal is attenuated by -70 dB. A commercial cryogenic circulator provides signal routing for the reflection measurement. On the \unit{1.6}{\kelvin}, \unit{4}{\kelvin} plate respectively, the signal is amplified by a commercial high electron mobility amplifier.
A combination of home-made and commercial low pass filters provides additional shielding from frequencies above \unit{8.2}{\giga\hertz}.
Fig.~\ref{supfig:1}c shows two closed waveguide sample holders attached to the copper rod, which provides thermalization and is used to mount the assembly on the mixing chamber plate of the cryostat. The sample holders are enclosed by a barrel made from a copper/aluminum sandwich, which we fix on a cap mounted on the central rod (schematically indicated). For additional magnetic shielding, we surround the copper/aluminum barrel by a $\mu$-metal shield (see Fig.~\ref{supfig:1}c, right picture).  

\subsection*{Evaluation of QP burst events}

We monitor each resonator for $\unit{45}{\minute}$ per readout power (cf. Fig.~3a in the main text) and vary the readout power applied to the sample holder input port from -120\,dBm to -165\,dBm (A), and -120\,dBm to -155\,dBm (B, C) in -5\,dBm steps. Table~\ref{suptab:2} shows how many events we identified at every readout power. Due to decreased signal to noise ratio, we identify less events at lower readout powers since events with smaller amplitude are not visible above the noise floor.

\begin{table}[!htb]
\begin{center}
\caption{\label{suptab:2} Number of individual events identified at every readout power during a total measurement time of \unit{45}{\minute} (cf. Fig.~3a in the main text). First, we identify traces with a QP burst event automatically, followed by a manual check of all identified traces.}
\newcolumntype{C}{>{\centering\arraybackslash}p{4em}}
\begin{tabular}{*{5}{C}}%|C|C|C|C|C|C|C|C|}
\hline \hline
\multicolumn{2}{c}{readout power} & \multicolumn{3}{c}{number of identified events} \\
\multicolumn{2}{c}{(dBm)} & A & B & C \\ \hline
\multicolumn{2}{c}{-120} & 103 & 93 & 72 \\
\multicolumn{2}{c}{-125} & 84 & 87 & 56 \\
\multicolumn{2}{c}{-130} & 75 & 79 & 54 \\
\multicolumn{2}{c}{-135} & 58 & 79 & 63 \\
\multicolumn{2}{c}{-140} & 72 & 81 & 62 \\
\multicolumn{2}{c}{-145} & 54 & 89 & 56 \\
\multicolumn{2}{c}{-150} & 59 & 62 & 56 \\
\multicolumn{2}{c}{-155} & 47 & 23 & 30 \\
\multicolumn{2}{c}{-160} & 48 & - & - \\
\multicolumn{2}{c}{-165} & 39 & - & - \\ \hline \hline
\end{tabular}

\end{center}
\end{table}

\begin{figure*}[!h]
\centering
\includegraphics[scale=0.9]{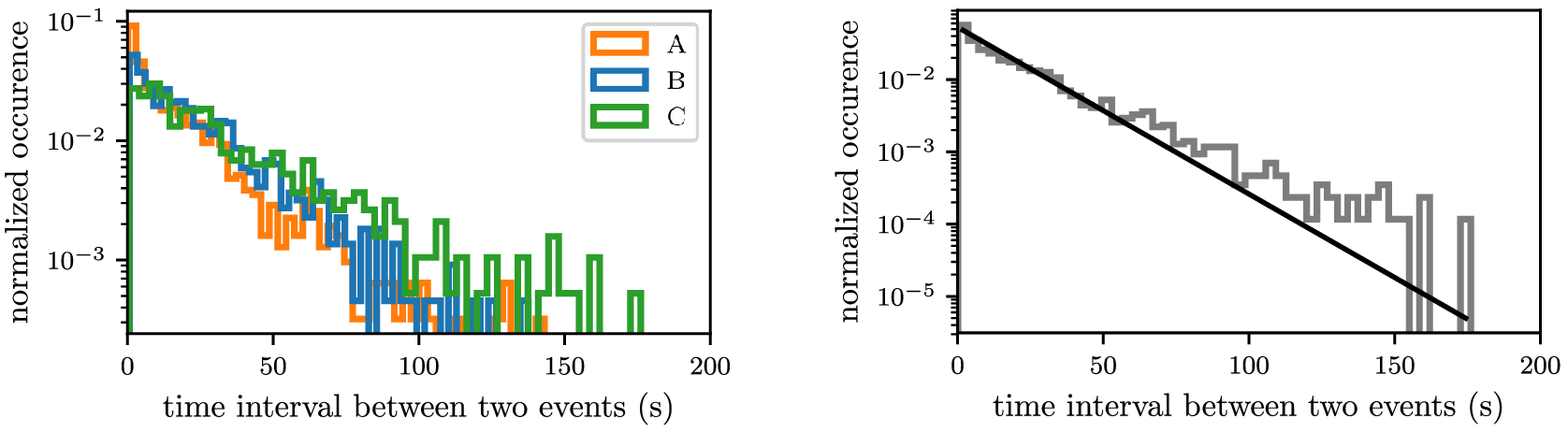}
\caption{\label{supfig:2a} Histograms of time intervals between two successive QP bursts. The left panel shows the histogram for resonators A, B, and C in log - lin scale. In the right panel we show the histogram of the cumulated data (grey) and a fit to an exponential distribution, from which we extract an average time of \unit{19}{\second} between two QP bursts.}
\end{figure*}

Figure~\ref{supfig:2a} shows histograms of the time between two successive QP bursts for resonators A, B, and C (left panel). The right panel shows the histogram of the cumulated data with a fit to an exponential probability distribution, $P(x) = \lambda \exp(-\lambda x)$, which describes events occurring continuously and independently at a constant average rate $\lambda$. From the fit we extract an average time between events of \unit{19}{\second}.

\begin{figure*}[htb]
\centering
\includegraphics[scale=1]{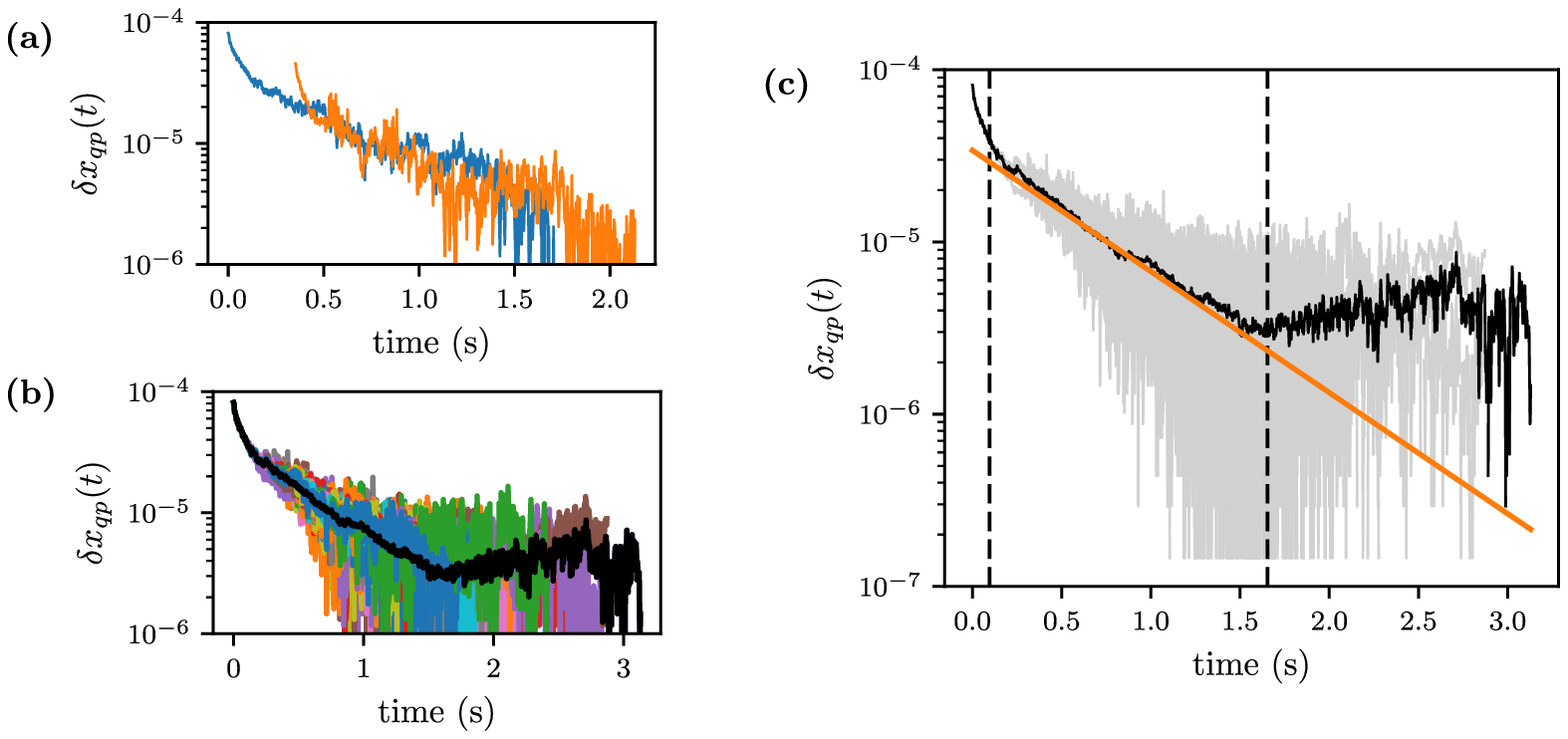}
\caption{\label{supfig:2} Fit procedure for the extraction of $\tau_{\mathrm{ss}}$. \textbf{(a)} In the initial step, we shift the events with respect to a reference event (blue line) to generate maximal overlap between the exponential tails of the events. \textbf{(b)} After the shifting, we remove the initial, steep decay of the curves and generate a mean event for comparison (black line). \textbf{(c)} We identify the exponential part of the dataset consisting of the shifted events with the help of the mean event (black line). For the final fit of $\tau_{\mathrm{ss}}$ we use all points in the previously identified exponential part of the dataset (black dashed lines). The noise of the data increases towards higher times because less data points are averaged.}
\end{figure*}

Figure~\ref{supfig:2} shows an overview of the fitting process for $\tau_{\mathrm{ss}}$. In a first step, we shift the individual events with respect to a reference event such, that all tails overlap (see Fig.~\ref{supfig:2}a). This amounts to a rescaling of the amplitudes of the exponential tails which does not alter their decay rate. The initial, steep recombination process shows different behavior specific to individual events, possibly due to varying origin of the QPs, varying diffusion processes, and fluctuations in the background QP density. After this initial phase however, the relaxation processes follows an exponential tail with universal slope for events at a given readout power. Following the shifting, we remove the initial steep part of all but the reference event's trace and plot the mean event for comparison (see Fig.~\ref{supfig:2}b). Finally, we identify the exponential part of the shifted events and use all data points in the interval to fit a line to the data in logarithmic-linear scale to extract the time constant of the exponential decay. 

\subsection*{Amplitude data of resonator C}

\begin{figure*}[!htb]
\centering
\includegraphics[scale=1]{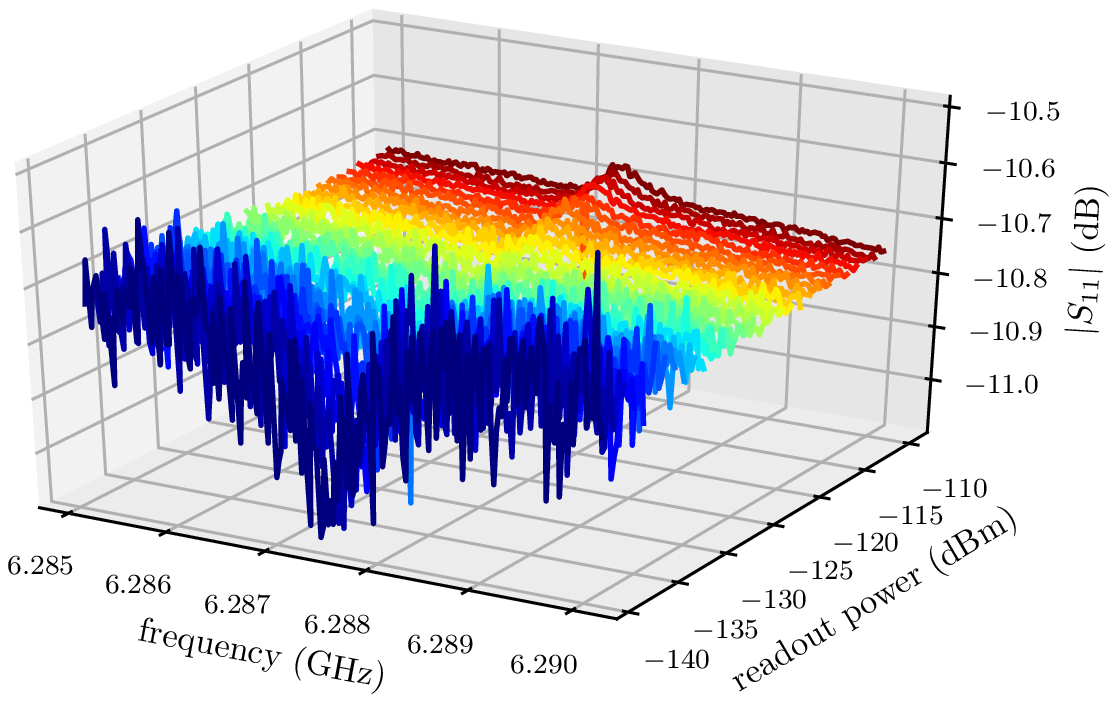}
\caption{\label{supfig:3} Amplitude data of resonator C as a function of the applied readout power. The amplitude response gradually changes from a slight peak of 0.1 dB at high readout powers, to a dip. This behavior could, among other reasons, result from imperfections in the impedance matching of the measurement setup or local flux trapping.}
\end{figure*}

In Fig.~2b in the main text, we only show fitted values of $Q_{\mathrm{i}}$ for resonators A and B. This is due to irregular behavior of the measured amplitude data of resonator C. Figure~\ref{supfig:3} shows how the amplitude signal gradually changes from a peak of approximately 0.1 dB to a dip at lower readout powers for a dataset recorded during cooldown \#3. Datasets from cooldowns \#1 and \#2 also show this behavior with a slight peak at high readout powers. The behavior of resonator C  could be caused, among other reasons, by imperfections in the impedance matching of the measurement setup or local flux trapping. 
For the estimation of the photon number $\bar{n}$ from the applied readout power in resonator C we approximate the total quality factor to be completely dominated by the coupling quality factor. We base this on results of $Q_{\mathrm{i}}$ of resonators A and B, and measurements of the phase response of resonator C, which shows a full $2\pi$ roll-off over the entire measured power range.

\subsection*{Internal quality factor $Q_{\mathrm{i}}$ for various IF bandwidth}

All resonators show QP producing impacts every $\sim \unit{20}{\second}$. In an averaged resonator spectrum frequency jumps could lead to a smaller depth of the observed dip in the amplitude signal of the resonator. Since we characterize our resonators by performing reflection measurements a reduced dip corresponds to smaller internal losses and an averaged spectrum would therefore yield higher internal quality factors than the true value. 
Furthermore, longer averaging is necessary to record data that can still be fitted at low readout powers. In order to quote a reliable number for the internal quality factor, we repeatedly measure the complex scattering parameters of the resonators with increasing averaging times, i.e. decreasing intermediate frequency bandwidth (IF BW). 
Figure~\ref{supfig:2} shows the internal quality factor $Q_{\mathrm{i}}$ as a function of the average number of photons circulating in the resonator for IF BW from $\unit{1000}{\hertz}$ to $\unit{10}{\hertz}$ corresponding to an increase in the averaging time by two orders of magnitude. As can be seen in the plot, all curves are in agreement. Expectedly, data measured with less averaging can only be fitted at higher readout powers. 
However, data taken with \unit{10}{\hertz} IF BW and \unit{50}{\hertz} IF BW are evaluated well into the single photon limit and the extracted numbers for $Q_{\mathrm{i}}$ are in good agreement. Therefore, we conclude that the internal quality factors quoted in the main text are achieved due to low losses in GrAl resonators and do not result from a decreased depth of the dip in amplitude data due to averaging.

\begin{figure*}[!htb]
\centering
\includegraphics[scale=1]{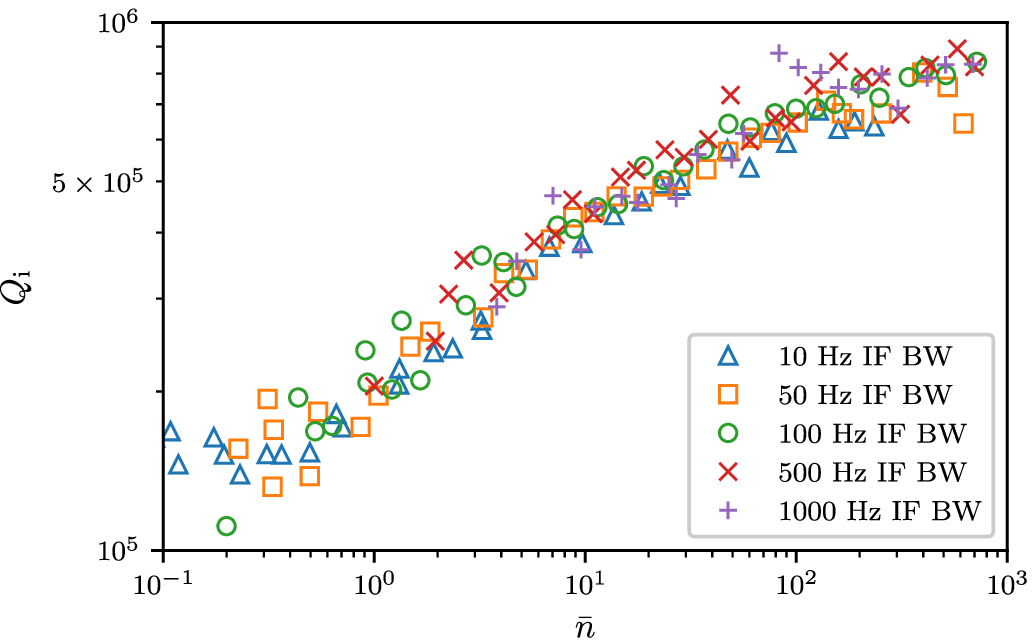}
\caption{\label{supfig:4} Internal quality factor $Q_\mathrm{i}$ as a function of the average photon number $\bar{n}$ for different averaging times. Results from fits to data with an averaging range spanning two orders of magnitude are in good agreement. Therefore, we conclude that the fit results of $Q_\mathrm{i}$ are not artificially increased by an averaging out of the dip due to the fluctuations of the resonant frequency, which could possibly be caused by QP bursts.}

\end{figure*}

\FloatBarrier

\subsection*{Phenomenological QP dynamics model}

In the presence of disorder, there are spatial variations of the superconducting order parameter. As discussed in Ref. \cite{bespalov_theoretical_2016}, QPs can be localized at these spatial variations, which induce a small subgap tail in the density of states. Here, we develop a phenomenological model similar to that introduced by Rothwarf and Taylor \cite{rothwarf_measurement_1967} that accounts for recombination of QPs, their generation, and transitions between localized and free (mobile) QPs. 

Let us indicate with $x_m$ and $x_l$ the normalized density of mobile and localized QPs, respectively ($x_m$ then is what is usually denoted with $x_\mathrm{qp}$). The time evolution of these normalized densities is assumed to be governed by the following set of coupled, non-linear differential equations:
\bea
\dot{x}_m & = & - \Gamma_{mm} x_m^2 - \Gamma_{ml} x_m x_l - \Gamma_\mathrm{loc} x_m + \Gamma_\mathrm{ex} x_l + g_m \label{xmeq}, \\
\dot{x}_l & = & - \Gamma_{ll} x_l^2 - \Gamma_{ml} x_m x_l + \Gamma_\mathrm{loc} x_m - \Gamma_\mathrm{ex} x_l + g_l \label{xleq}.
\eea
In both equations, the first two terms on the right hand side account for recombination between QPs of the same kind, with rates $\Gamma_{mm}$ or $\Gamma_{ll}$, and between different types of QPs, with rate $\Gamma_{ml}$. The terms proportional to the rate $\Gamma_\mathrm{loc}$ describe decay of QPs from above to below the gap -- that is, the localization of mobile QPs -- and the rate $\Gamma_\mathrm{ex}$ the opposite excitation process. Finally, $g_m$ and $g_l$ are the generation rates of mobile and localized QPs, respectively. Note that assuming a constant density of localized QPs, with the identifications
\be
\Gamma_{ml}x_l + \Gamma_\mathrm{loc} \to s \, , \qquad \Gamma_\mathrm{ex}x_l + g_m \to g \, , \qquad \Gamma_{mm} \to r \, ,
\ee
\eref{xmeq} reduces to the phenomenological equation used in e.g. Ref. \cite{Wang_NatComm2014}.

Since at long times after an event generates QPs, the deviations from the steady-state are small, we can linearize the above equations by separating the small, time-dependent deviation from the steady state density:
\be
x_m(t) = \bar{x}_m + \delta x_m(t) \, , \qquad x_l(t) = \bar{x}_l + \delta x_l(t).
\ee
Keeping terms up to the first order we find
\bea
0 & = & - \Gamma_{mm} \bar{x}_m^2 - \Gamma_{ml} \bar{x}_m \bar{x}_l - \Gamma_\mathrm{loc} \bar{x}_m + \Gamma_\mathrm{ex} \bar{x}_l + g_m \label{xmbeqf},\\
0 & = & - \Gamma_{ll} \bar{x}_l^2 - \Gamma_{ml} \bar{x}_m \bar{x}_l + \Gamma_\mathrm{loc} \bar{x}_m - \Gamma_\mathrm{ex} \bar{x}_l + g_l \label{xlbeqf},
\eea
and
\bea
\delta\dot{x}_m & = & - 2\Gamma_{mm} \bar{x}_m \delta x_m - \Gamma_{ml} \bar{x}_m \delta x_l  - \Gamma_{ml} \bar{x}_l \delta x_m - \Gamma_\mathrm{loc} \delta x_m + \Gamma_\mathrm{ex} \delta x_l, \label{xmdeq}\\
\delta\dot{x}_l & = & - 2\Gamma_{ll} \bar{x}_l \delta x_l- \Gamma_{ml} \bar{x}_m \delta x_l  - \Gamma_{ml} \bar{x}_l \delta x_m + \Gamma_\mathrm{loc} \delta x_m - \Gamma_\mathrm{ex} \delta x_l. \label{xldeq}
\eea

The general model in \esref{xmeq}-\rref{xleq} can be simplified by considering the microscopic origin of the various rates. At low temperatures and small QP density, recombination and scattering rates are determined by electron-phonon interaction. For a zero-temperature phonon bath, the generation rates due to phonons would be exactly zero, but QPs can be also generated by photons and/or other elementary particles (e.g., protons \cite{bespalov_theoretical_2016}) of sufficient energy; assuming this energy to be large compared to the gap, we can set $g_l=0$. Localization and recombination take place by phonon emission; based on Ref. \cite{kaplan_quasiparticle_1976}, at low temperatures, we expect $\Gamma_\mathrm{loc} > \Gamma_{mm}$, and since $x_m \ll 1$, we neglect the term proportional to $\Gamma_{mm}$ (in fact, this approximation is applicable under the much less stringent condition $\Gamma_{mm}x_m \ll \Gamma_\mathrm{loc}$). Later on, we will consider excitation due to photons in the resonator by setting
\be
\Gamma_\mathrm{ex} = \Gamma_0 \bar{n},
\ee
while neglecting photon emission by QPs in comparison to phonon emission. This approximation should be valid so long as $\Gamma_0 \bar{n} \ll \Gamma_\mathrm{loc}$. 

\subsection*{Steady-state}

With the simplifications, \esref{xmbeqf}-\rref{xlbeqf} become
\bea
0 & = &  - \Gamma_{ml} \bar{x}_m \bar{x}_l - \Gamma_\mathrm{loc} \bar{x}_m + \Gamma_\mathrm{ex} \bar{x}_l + g_m, \label{xmbeqs}\\
0 & = & - \Gamma_{ll} \bar{x}_l^2 - \Gamma_{ml} \bar{x}_m \bar{x}_l + \Gamma_\mathrm{loc} \bar{x}_m - \Gamma_\mathrm{ex} \bar{x}_l. \label{xlbeqs}
\eea
Solving the last equation for $\bar{x}_l$ in terms of $\bar{x}_m$, we find
\be
\bar{x}_l = \frac{\sqrt{\left(\Gamma_{ml}\bar{x}_m+\Gamma_\mathrm{ex}\right)^2+4\Gamma_{ll}\Gamma_\mathrm{loc}\bar{x}_m}-\left(\Gamma_{ml}\bar{x}_m+\Gamma_\mathrm{ex}\right)}{2\Gamma_{ll}}.
\ee
With the further assumption
\be\label{fa}
\Gamma_{ll}\Gamma_\mathrm{loc} \ll \Gamma_{ml}^2\bar{x}_m,
\ee
we have
\be\label{xlbf1}
\bar{x}_l \simeq \frac{\Gamma_\mathrm{loc}\bar{x}_m}{\Gamma_{ml}\bar{x}_m+\Gamma_\mathrm{ex}}.
\ee
Substituting this expression into \eref{xmbeqs}, we get
\be\label{xmbf}
\bar{x}_m \simeq \frac{g_m}{2\Gamma_\mathrm{loc}} \frac12 \left(1+\sqrt{1+8\frac{\Gamma_\mathrm{loc}\Gamma_\mathrm{ex}}{g_m\Gamma_{ml}}}\right)
= \frac{g_m}{2\Gamma_\mathrm{loc}} \left[1+ \frac12\left(\sqrt{1+4\gamma\bar{n}}-1\right)\right] \, , \qquad \gamma = 2 \frac{\Gamma_\mathrm{loc}\Gamma_0}{g_m\Gamma_{ml}}.
\ee
Then we can rewrite \eref{xlbf1} as
\be\label{xlbf2}
\bar{x}_l = \frac{\Gamma_\mathrm{loc}}{\Gamma_{ml}}\frac{1}{1+\frac{\gamma\bar{n}}{1+ \frac12\left(\sqrt{1+4\gamma\bar{n}}-1\right)}}.
\ee

\subsection*{Decay rate}

From \esref{xmdeq}-\rref{xldeq}, we generically obtain two decay rates $\lambda_\pm$, a fast one (+) and a slow one (-):
\be
\lambda_\pm = \frac12\left[a+b \pm \sqrt{(a-b)^2+4c}\right],
\ee
with
\be\label{abc}
a = 2\Gamma_{ll}\bar{x}_l + \Gamma_{ml}\bar{x}_m + \Gamma_\mathrm{ex}\, , \qquad b = 2\Gamma_{mm}\bar{x}_m + \Gamma_{ml}\bar{x}_l + \Gamma_\mathrm{loc} \, , \qquad c = \left(\Gamma_\mathrm{loc}-\Gamma_{ml}\bar{x}_l\right)\left(\Gamma_\mathrm{ex}-\Gamma_{ml}\bar{x}_m\right).
\ee

A limiting regime is when $|c| \ll (a-b)^2$, in which case the two decay rates are approximately given by $a$ and $b$. As we discuss in the next section, we expect in practice $a < b$, so $a$ is the slow mode. Then under the assumption in \eref{fa}, which leads to \eref{xlbf1}, we can neglect the first term in the definition of $a$ with respect to the second term and we get
\be\label{lmf}
\lambda_- \simeq a \simeq \Gamma_{ml}\bar{x}_m + \Gamma_\mathrm{ex} \simeq \Gamma_0 \left\{ \frac{1}{\gamma} \left[1+ \frac12\left(\sqrt{1+4\gamma\bar{n}}-1\right)\right] + \bar{n}\right\}.
\ee

\subsection*{Comparing to experiments}

The inverse quality factor can be generically written as the sum of the inverse of the quality factor at zero average photons $Q_0$ (due to both QPs present in the absence of photons and other loss mechanisms, such as dielectric losses) plus a photon number dependent part due to the change in the number of QPs. In principle, both localized and mobile QPs can contribute to the losses in a way proportional to their normalized density times coupling strength times final density of states. Since the change in quality factor is less than one order of magnitude, we expect the change in normalized density to also be small, and to compare localized and mobile QPs we can use their zero-photon values.
We also expect $\Gamma_{ml}<\Gamma_{mm}$, since the spatial overlap between localized and mobile QPs cannot be larger than that between mobile QPs. Then, the simplifying assumption $\Gamma_{mm} \bar{x}_m \ll \Gamma_\mathrm{loc}$ implies $\bar{x}_l > \bar{x}_m$. The final density of states is likely larger for localized QPs than for mobile QPs, for example if the former go from below the gap to just above the gap, while the latter always end at a higher energy above the gap, where the density of states is smaller. Therefore, if the coupling strengths to photons are similar, the localized QPs give a larger contribution to the losses (i.e., inverse quality factor), and using \eref{xlbf2} we can write
\be
\frac{1}{Q_i} = \frac{1}{Q_0} + \beta \left[ \frac{1}{1+\frac{\gamma\bar{n}}{1+ \frac12\left(\sqrt{1+4\gamma\bar{n}}-1\right)}} - 1\right],
\ee
where factor $\beta$ is proportional to $\Gamma_\mathrm{loc}/\Gamma_{ml}$ but also accounts for coupling strength and final density of states. Note that the formula cannot be extrapolated to very large $\bar{n}$ for two reasons: according to Eq.~\eqref{xmbf}, the density of mobile QPs increases with $\bar{n}$; therefore their contribution to the quality factor can become relevant, and also neglecting the recombination between mobile QPs with respect to their localization will not hold anymore.

For the inverse of the decay time, we can proceed in a similar way and sum a photon-dependent part [from \eref{lmf}] to a residual decay rate $\Gamma_r$, which includes both the zero-photon decay process which is part of the model, as well as other mechanisms not explicitly accounted for:
\be
\frac{1}{\tau_\mathrm{ss}} = \Gamma_{r} + \Gamma_0 \left[\bar{n} + \frac{1}{2\gamma}\left(\sqrt{1+4\gamma\bar{n}}-1\right)\right].
\ee
Note that, since we have included the zero-photon contribution of Eq.~\eqref{lmf} into $\Gamma_r$, we should always have $\Gamma_r \ge \Gamma_0/\gamma$. 

Finally, let us check for consistency of assumptions: let us use $\gamma\sim 1$ and $\Gamma_0 \sim 10^{-2}$ s$^{-1}$, see Fig. 4 in the main text. From Ref. \cite{Wang_NatComm2014} we have $\bar{x}_m \sim 10^{-6}$, $g_m \sim 10^{-4}\,$s$^{-1}$ and $\Gamma_{mm} \sim 10^7\,$s$^{-1}$. Using these values in \eref{xmbf} (with $\bar{n}=0$) we estimate $\Gamma_\mathrm{loc}\sim 10^2\,$s$^{-1}$, and using the definition of $\gamma$ in the same equation we also estimate $\Gamma_{ml}\sim 10^4\,$s$^{-1}$. Therefore, we find that indeed $\Gamma_{ml} < \Gamma_{mm}$, and also $\Gamma_{mm} \bar{x}_m \ll \Gamma_\mathrm{loc}$. Even for $\bar{n} \sim 10^2$ we have $\Gamma_0\bar{n}\ll\Gamma_\mathrm{loc}$. With these estimates, the assumption in \eref{fa} becomes $\Gamma_{ll} \ll 1\,$s$^{-1}$, but to our knowledge there are no experimental data on this rate. Keeping this assumption, we can also estimate the quantities in \eref{abc}: $a \sim \Gamma_0 \left(1/\gamma + \bar{n}\right) \lesssim 1$~s$^{-1}$ (for $\bar{n}\lesssim 10^2$), $b > \Gamma_\mathrm{loc} \sim 10^2$~s$^{-1}$, and $|c|\lesssim \Gamma_0 \Gamma_\mathrm{loc} \bar{n} \lesssim 10^2$ s$^{-2}$ (for $\bar{n}\lesssim 10^2$). These estimates verify the assumptions $a<b$ and $(a-b)^2 \gg |c|$.

\end{document}